\documentclass[letter,11pt]{article}
\usepackage{amsgen,amsmath,amstext,amsbsy,amsopn,amssymb}
\usepackage{times}
\usepackage{graphicx}
\usepackage[margin=1in]{geometry}
\usepackage{setspace}
\usepackage[font={footnotesize}]{caption}
\usepackage{enumitem}
\usepackage{blkarray}
\usepackage{natbib}
\usepackage{multirow}
\usepackage{booktabs}

\newcommand{\reals}{\mathbb{R}}

\newcommand{\x}{{\mathbf{x}}}

\newcommand{\z}{{\mathbf{z}}}

\newcommand{\btau}{\boldsymbol{\tau}}

\newcommand{\bt}[1]{\mbox{\boldmath \mbox{$#1$}}}
\DeclareGraphicsExtensions{.pdf,.png,.jpg}

\begin{document}
\title{\textbf{Model-based Dashboards for Customer Analytics}}
\author{Ryan Dew\footnote{\noindent Ryan Dew (email: \texttt{ryan.dew@columbia.edu}) is a doctoral student in Marketing at Columbia Business School, Columbia University.}~ and Asim Ansari\footnote{Asim Ansari (email: \texttt{maa48@columbia.edu}) is the William T. Dillard Professor of Marketing at Columbia Business School, Columbia University.} \\ Columbia University}

\pagenumbering{gobble}

\maketitle

\begin{abstract}
\noindent Automating the customer analytics process is crucial for companies that manage distinct customer bases. In such data-rich and dynamic environments, visualization plays a key role in understanding events of interest. These ideas have led to the popularity of analytics dashboards, yet academic research has paid scant attention to these managerial needs. We develop a probabilistic, nonparametric framework for understanding and predicting individual-level spending using Gaussian process priors over latent functions that describe customer spending along calendar time, interpurchase time, and customer lifetime dimensions. These curves form a dashboard that provides a visual model-based representation of purchasing dynamics that is easily comprehensible. The  model flexibly and automatically captures the form and duration of the impact of events that influence spend propensity, even when such events are unknown a-priori. We illustrate the use of our Gaussian Process Propensity Model (GPPM) on data from two popular mobile games. We show that the GPPM generalizes hazard and buy-till-you-die models by incorporating calendar time dynamics while simultaneously accounting for recency and lifetime effects. It therefore provides insights about spending propensity beyond those available from these models. Finally, we show that the GPPM outperforms these benchmarks both in fitting and forecasting real and simulated spend data. 
\end{abstract}

\pagenumbering{arabic}

\section{Introduction}

Marketers in multi-product companies face the daunting task of monitoring and understanding the ebb and flow of aggregate sales for many different products and customer bases. The dynamic pattern of purchases for any given product stems from both the impact of myriad managerial actions on customers, and from the more natural underlying stochastic dynamics of customer purchasing activities and lifetime events. The marketing analyst and the personnel who oversee multiple products are often not fully cognizant of the entire set of underlying events that can impact sales of a given product, especially if product decisions are made within the product team. Modern day managers often seek to manage such challenges caused by multi-product contexts and information asymmetry via decision support tools, such as marketing dashboards, that summarize  key customer base metrics and visually portray dynamic patterns across  products and customer bases. 

Most marketing dashboards are based on a combination of exploratory data analysis (EDA) tools that mine datasets for patterns, time series methods~\citep{Hanssens2001}, and predictive data mining methods, such as bagging and boosting~\citep{Neslin2006} that predict future activity. While these tools are indeed popular in decision support~\citep[for an overview, see][]{Pauwels2009}, a more principled, model-based approach can be useful in developing a consistent understanding of customer dynamics across different contexts. In this paper, we develop and illustrate an approach that leverages Bayesian nonparametric methods to probabilistically model dynamic patterns in purchasing. This flexible and robust framework is based on Gaussian processes, and integrates data at multiple time scales and across levels of aggregation to facilitate model-based visualization of spending dynamics across multiple products, in an automatic fashion.    

The need for such model-based dashboards is paramount in our partner company, a large American video game company. The company develops and publishes games that span a variety of platforms and genres, each managed by its own game team that is responsible for both product development and for product changes after launch. This latter operation is especially relevant in the mobile video game space, where the primary approach of the company is to launch a game once, and then release new content for the game over time. This strategy, facilitated by the fact that mobile games are inherently tethered to the internet, is quite different from the traditional strategy of releasing new content via game sequels. It also poses a unique challenge for marketing analysts and managers who need to simultaneously analyze many of these games: information about changes to a particular game does not necessarily filter automatically to the central marketing analyst, or to layers of management in charge of multiple games and genres. Thus, covariates of interest may not be fully observed or shared. At the same time, managers need to quickly assess in a panoramic fashion the events and dynamics associated with each game, and be able to filter the true signal from the noise in spending patterns. 

While this problem may seem specific to digital industries, it is also prevalent in traditional product markets, as sales of a given product-line can be affected by events that are outside the scope of awareness of marketing managers. New product launches from competitors, disruptions in the supply chain, and real world events like holidays may accelerate or hinder purchasing and information asymmetries may abound, as most products have different management teams. Yet still, it is critical for top-management to detect, cope with, and understand the impact of such managerial action and to grasp the importance of patterns in customer base activities.

Model-based understanding of customer-level data and of the aggregate customer base dynamics inherent in such data is central to much research in marketing. Prominent contributions  have come from a long and distinguished tradition of research on customer base analysis that relies on stochastic characterizations of customer events based on interpurchase times and customer lifetimes, in contractual and non-contractual settings  (\citealp{Chatfield1973,Morrison1981,Schmittlein1987,Fader2005},~\citealp{Fader2005a},~\citealp{Fader2010,Ascarza2013,Schweidel2013}). These models characterize the heterogeneity across customers and predict short-term purchasing activity, but generally rely on only interpurchase times for estimation. One notable exception to this sole reliance is~\citealp{Schweidel2013}, which nicely incorporates the calendar time impact of direct marketing activities into a customer base modeling framework. Managers think and act in calendar time and focus more on aggregate dynamics, and thus, a model-based approach that is capable of synthesizing calendar time activity and customer-base events with these recency and lifetime models is important. However, an ideal framework must also be able to detect and accommodate unknown or unobserved shocks to spend propensities, in addition to observed and known direct marketing activities.

In an approach that is novel to the marketing literature, we fuse together latent functions that operate along three dimensions---calendar time, interpurchase time (recency), and customer lifetime---using Bayesian nonparametric Gaussian process priors, to bring calendar time insights into the customer base analysis framework. We use these latent functions in tandem with standard controls for observed and unobserved heterogeneity within a discrete hazard framework to dynamically model customer purchase propensities. We term the resulting model the Gaussian Process Propensity Model (GPPM). Our reliance on a nonparametric method means that the shapes of these latent functions are automatically inferred from the data, thus providing the flexibility to robustly adapt to different settings, and to capture time-varying effects, even in contexts where all the information about inputs may not be available. The use of latent functions also allows us to develop visual representations of the latent dynamics, thus facilitating easy comparison across products. While Bayesian nonparametrics have been successfully applied to marketing problems~\citep[e.g.][]{Wedel2004, Kim2007, Rossi2013, Li2014}, our paper is the first to our knowledge to take advantage of the powerful Gaussian process prior methodology.

In this paper, we describe in detail how Gaussian processes can be used to specify latent dynamics and show what insights can be gleaned from our model. We illustrate our approach on data from two mobile video games owned by our partner company. These games span different customer bases and we show how the visualizations generated from our approach can facilitate managerial understanding of the key dynamics within each customer base. We also compare the GPPM to benchmark probability models, in particular the BGNBD model (Fader, Hardie and Lee, 2005) and the hierarchical log-logistic hazard model. We show that the GPPM is a natural generalization of these models, as it captures calendar time effects in addition to recency and lifetime effects. In this way, we isolate the gains that come from modeling multiple time scales. We also demonstrate the superiority of our framework using both real and simulated datasets. In particular, we show that the GPPM can provide a unique set of insights regarding both the static and dynamic patterns of behavior underlying customer purchase data. We also show that the GPPM predicts better than the benchmark models, both in-sample and out-of-sample. 

The rest of the paper is organized as follows. The next section describes the GPPM framework, discusses our model formulation, and highlights the benefits of the Bayesian nonparametric approach. Section 3 describes the application of our model to the data from two games, presents the results, and discusses the insights that can be gleaned from the GPPM framework. Section 4 focuses on the model's forecasting capacity and presents model comparison results. Finally, Section 5 concludes with a summary, limitations, and directions for future research.

\section{Model}

In the GPPM framework, latent propensity functions are modeled by utilizing the natural variability in spend incidence data along three time dimensions: calendar time, interpurchase time, and customer lifetime. Our focus on modeling spending incidence is consistent with the majority of the literature on customer base analysis, and also fits nicely with our application area, where there is minimal variability in spend amount, but considerable variability in spend frequency. These propensity functions are nested in a discrete time hazard framework, as most customer-level data are available at a discrete level of aggregation. In our application, the data are available at a daily level.

The observations in our data consist of a binary indicator $y_{it}$ that specifies whether customer $i$ made a purchase at calendar time $t$, and a corresponding triple $(t,r_{it},\ell_{it})$ for calendar time, recency, and customer lifetime respectively. Recency here refers to interpurchase time, or the time since the customer last purchased, while customer lifetime refers to the time since the customer first purchased. Depending on the context, a vector of time varying covariates $\x_{it}$ and a vector $\z_{i}$ of demographics or fixed traits of interest, such as the customer acquisition channel and acquisition date, may also be available. The probability of customer $i$ making a purchase at time $t$ is modeled in the GPPM as
\begin{equation}
\mbox{Pr}(y_{it}=1) = \mathrm{logit}^{-1}\left[\alpha(t,r_{it},\ell_{it})+\x_{it}^{\prime}\bt{\beta}+\z_{i}^{\prime}\bt{\gamma} + \delta_i\right],
\end{equation}
where, $\mathrm{logit}^{-1}(\cdot)$ denotes the canonical inverse logit link function:
\begin{equation}\mathrm{logit}^{-1}(x) = \frac{1}{1+\exp(-x)}. \end{equation}
We see in Equation 1 that the spending rate is driven by a time-varying component $\alpha(.)$ and a variety of effects. The random effect $\delta_i$ captures unobserved customer heterogeneity in spending propensities. This setup makes the implicit assumption that the dynamics are aggregate trajectories---that is, all customers follow the same dynamic process---while maintaining individual heterogeneity via the random effect $\delta_i$ and by using other observed individual-specific variables.

The heart of our framework involves the specification of the time-varying component $\alpha(t,r_{it},\ell_{it})$. We treat $\alpha(.)$ as a latent function and model it nonparametrically using Gaussian process priors~\citep{Rasmussen2006,Roberts2013}. The nonparametric approach models random functions flexibly and allows us to  automatically accommodate different patterns of spend dynamics that may underlie a given customer base. These dynamics operate along  all three of our time dimensions, and at different time scales within a dimension. Such dynamics may include smooth long-run trends and short-term patterns, as well as cyclic variation, and are automatically inferred from data. To allow such rich structure, we use an additive mixture of unidimensional GPs to specify and estimate the multivariate function $\alpha(t,r_{it},\ell_{it})$. 

\subsection{Gaussian Process Priors}

To make clear the benefits of this approach, we first describe what GPs are, and how they can nonparametrically capture rich dynamic patterns in a Bayesian probability model. Suppose $f$ is a function that depends on a single dimensional input, $\tau \in \reals$ (e.g., time), and let $\bt{\tau}=\{\tau_{1}, \tau_{2}, \ldots, \tau_{N}\}$ be a vector of $N$ of these scalar inputs. A Gaussian process is a distribution over unknown functions $f$, such that for any finite set of input values, the function values $f(\tau_{1})$, $f(\tau_{2})$, \ldots, $f(\tau_{N})$ are jointly distributed multivariate Gaussian. From a Bayesian perspective, a GP provides a natural mechanism for probabilistically specifying uncertainty over a function space, and therefore acts as a prior for unknown functions. A GP prior is completely specified by its mean function, 
$
\mbox{E}[f(\bt{\tau})]=m(\bt{\tau}),
$
and its covariance function,
$
\mbox{Cov}[f(\bt{\tau}), f(\bt{\tau}^{\prime})]=K(\bt{\tau}, \bt{\tau}^{\prime}).
$
This covariance matrix is composed from a covariance kernel function $k(\tau_{i}, \tau_{j})$ such that
\begin{equation}
K(\bt{\tau}, \bt{\tau})=
\begin{pmatrix}
k(\tau_{1},\tau_{1})  & k(\tau_{1},\tau_{2}) &  \dots & k(\tau_{1},\tau_{N}) \\
k(\tau_{2},\tau_{1})  & k(\tau_{2},\tau_{2}) &  \dots & k(\tau_{2},\tau_{N}) \\
\vdots & \vdots & \ddots & \vdots \\
k(\tau_{N},\tau_{1})  & k(\tau_{N},\tau_{2}) &  \dots & k(\tau_{N},\tau_{N}) 
\end{pmatrix}.
\end{equation}
Given the mean function, kernel, and a vector of inputs $\btau$, the vector of function outputs associated with those inputs, $f(\btau) = \{ f(\tau_{1})$, $f(\tau_{2})$, \ldots, $f(\tau_{N})$ \}, is distributed multivariate Gaussian, 
\begin{equation}
f(\bt{\tau}) \sim {\cal N}(m(\bt{\tau}),  K(\bt{\tau}, \bt{\tau})).
\end{equation}
Understanding a GP as an infinite multivariate normal, and a function estimated using a GP prior as a finite multivariate normal projection of that infinite dimensional object makes it very simple to comprehend how function interpolation and extrapolation work in this framework. Conditioned on an estimate for the mean function and kernel, predicting new values for the latent function $f$, given some new input points $\bt{\tau}^*$, is equivalent to predicting new values conditionally from a multivariate normal. Specifically, the joint distribution of the old and new function values is given by
\begin{equation} \left[\begin{array}{c}
f(\bt{\tau}) \\ 
f(\bt{\tau}^*)
\end{array}\right]  \sim \mathcal{N} \left(\left[\begin{array}{c}
m(\bt{\tau}) \\ 
m(\bt{\tau}^*)
\end{array}\right], \left[ \begin{array}{cc}
K(\bt{\tau},\bt{\tau}) & K(\bt{\tau},\bt{\tau}^*) \\ 
K(\bt{\tau}^*,\bt{\tau}) & K(\bt{\tau}^*,\bt{\tau}^*)
\end{array} \right]  \right), \end{equation}
and hence the conditional distribution of the new inputs can be written as
\begin{align}
f(\bt{\tau}^*)  \sim \mathcal{N}( & m(\bt{\tau}^*)+K(\bt{\tau}^*,\bt{\tau}) K(\bt{\tau},\bt{\tau})^{-1}[f(\bt{\tau})-m(\bt{\tau})], \nonumber \\
& K(\bt{\tau}^*,\bt{\tau}^*) - K(\bt{\tau}^*,\bt{\tau})K(\bt{\tau},\bt{\tau})^{-1}K(\bt{\tau},\btau^*))
\end{align} 
The mean function and the covariance kernel determine the nature of the functions that a GP generates. The choice of the kernel, in particular, allows us to model functions with different structures and levels of smoothness.  Both the mean function and the covariance kernel can depend on hyperparameters that can be inferred from the data. Often, a zero mean function is used, if no prior information exists about the expected shape of the functions.

\subsubsection{Kernels}
The kernel defines the fundamental structure of a GP, and thus also defines the latent function space of a GP prior. Many different forms of the kernel have been explored in the GP literature~\citep{Genton2001,Rasmussen2006,Duvenaud2013}. In this paper we use two types of kernels that are suitable for our context. The first is the squared exponential kernel defined by
\begin{equation} k_{\textrm{SE}}(\tau_i,\tau_j;\eta,\rho) = \eta^2 \exp\left\{-\frac{(\tau_i-\tau_j)^2}{\rho^2}\right\} \end{equation}
where the hyperparameter $\eta$ is called the amplitude and $\rho$ is called the characteristic length-scale or ``smoothness." The amplitude can be best explained by considering the case when $\tau_i=\tau_j\equiv\tau$. In this case, $k(\tau,\tau)=\eta^2$, which is the variance of the normal distribution at the fixed input $\tau$. More generally, $\eta^2$ captures variability around the mean function, and in the case of a flat mean function, represents the output scale of the function. The characteristic length-scale $\rho$ intuitively indicates how far apart two input points need to be for the corresponding outputs to be uncorrelated. Hence, a high value of $\rho$ corresponds to very smooth functions, while a small value of $\rho$ yields jagged, unpredictable functions.

We also use a strictly cyclic generalization of the squared exponential kernel, defined by
\begin{equation} k_{\textrm{C}}(\tau_i,\tau_j;\omega,\eta,\rho) = \eta^2 \exp \left\{ -\frac{\sin^2\left(\pi (\tau_i-\tau_j)^2 / \omega \right)}{\rho^2} \right\}. \end{equation}
This allows for strictly cyclic functions of cycle length $\omega$, again defined by an amplitude $\eta$ and a length-scale $\rho$. Note that this variability could, hypothetically, also be captured by the squared exponential kernel; the benefit of using the strictly cyclic kernel is that forecasts from this GP will always precisely mirror the estimated pattern. Hence, if there is predictable cyclic variability in the data, this would be captured both in and out-of-sample. The squared exponential kernel, on the other hand, captures only ideas of smoothness and amplitude, and therefore will not produce strictly cyclic behavior out-of-sample. We choose the above two kernels because they are simple to specify and have easily understandable parameters. Moreover, despite their simplicity, they are capable of generating a variety of functional forms, and as we will see, can capture key ideas of fundamental structure. 

The GP literature has used a variety of approaches, such as optimization, cross-validation, and model-based methods for estimating the hyperparameters.  We use a standard Bayesian framework and treat hyperparameters as unknowns that are jointly estimated along with the latent function values. The  GP specification in which the multivariate normal acts as a prior over function values  with a covariance specified via a small number of hyperparameters governing broad aspects of the fitted function enables a natural tradeoff between fit and complexity. Therefore, overfitting is generally not a problem in GPs (for details, see \citealp{Rasmussen2006}, Chapter 5). 

\subsubsection{GP Function Spaces}

\begin{figure}[p!]
\centering
\makebox[\textwidth][c]{\includegraphics[scale=0.75]{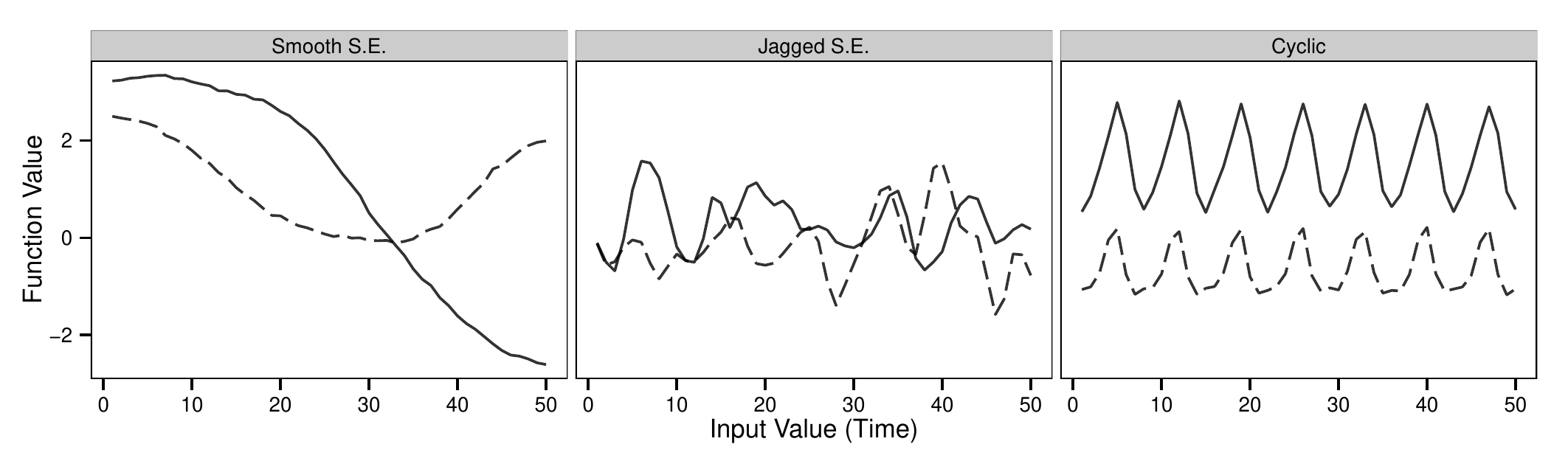}}
\caption{Random functions drawn from three Gaussian Processes with zero mean, $m(\tau)=0$. The left panel shows two draws from a GP with a squared exponential kernel ($\eta^2=3$ and $\rho^2=300$). The middle panel is based on a GP with a SE kernel with $\eta^2=0.5$ and $\rho^2=5$. The right panel uses a strictly cyclic kernel with $\eta^2=3$, $\rho^2=5$, and $c=7$.}
\label{fig:gpex1}
\end{figure}

We use Figure~\ref{fig:gpex1} to visually illustrate how the kernel and kernel hyperparameters of a GP influence the nature of the functions it generates. The figure shows functions drawn from three different Gaussian process priors, all of which have a zero mean function. The first two panels use the squared exponential kernel with different amplitude and smoothness hyperparameters. We see in the first panel that a GP with a larger amplitude parameter generates functions with a larger range of function values. Similarly, a larger smoothness or length-scale implies a higher covariance between function values over a larger range, resulting in a smoother function. The last panel displays a cyclic function drawn from the strictly cyclic kernel. 

\begin{figure}[p!]
\centering
\makebox[\textwidth][c]{\includegraphics[scale=0.75]{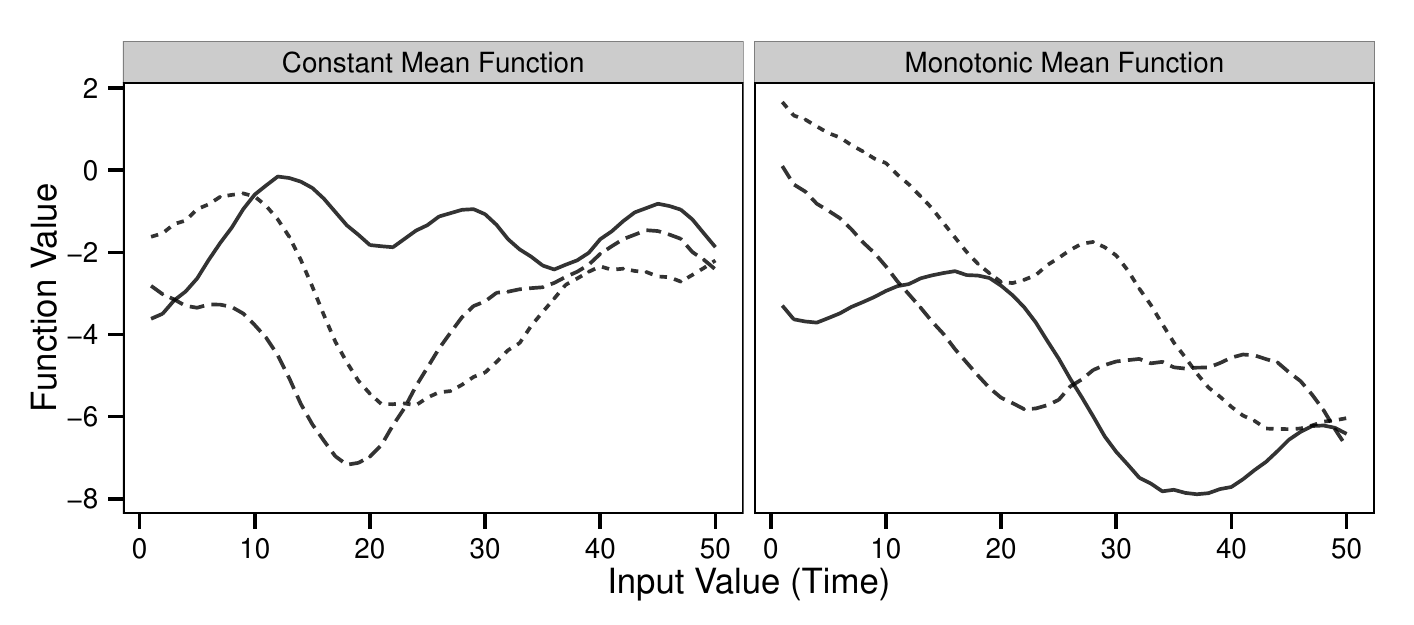}}
\caption{Random functions drawn from two GPs with $\eta^2=3$, $\rho^2=100$ and non-zero mean functions. The left panel uses a mean function, $m(\tau)=-3$. The right panel has $m(\tau)=-0.5(\tau-1)^{0.7}$, a monotonic decreasing function.}
\label{fig:gpex2}
\end{figure}

Figure~\ref{fig:gpex2} shows the impact of the mean function. The mean function at an input point primarily acts to shift the center of the function values at that point. Hence, for a  constant mean function of $m(\tau)=-3$, as in the left panel, the function values tend to be centered around -3. In contrast, the functions shown in the  right panel of Figure~\ref{fig:gpex2} come from a GP with a monotonically decreasing mean function and exhibit a downward trend, on average. Note, however, that a wide range of function shapes, including non-monotonic ones, are still possible even with this explicit mean function. In practice, the mean function for a GP prior encodes our best guess of the functional form and drives the posterior estimates of the function in areas where the input data are very sparse. In areas where the input data are dense, the prior guess is overwhelmed by the data.  When the hyperparameters in the mean function and the covariance kernel are estimated from the data, rather than imposed, then both the nature of the latent function space and the function values themselves are learned from the data. 

\subsubsection{Multidimensional GPs and Additivity}

In practice, we often care about estimating a latent multidimensional function, as in our application. Let $g(\tau^{(1)},\tau^{(2)},\ldots,\tau^{(D)})$ be a function from $\reals^D$ to $\reals$. Now, inputs are vectors of the form $(\tau_i^{(1)},\tau_i^{(2)},\ldots,\tau_i^{(D)}) \in \reals^D$, for $i=1,\ldots,N$, such that the set of all inputs is an $N \times D$ matrix. Just as before, $g$ can also be modeled via a GP prior. While there are many approaches for modeling multi-input functions with GPs, a simple yet powerful approach is to consider $g$ as a sum of single input functions, $g_1,g_2,\ldots,g_D$, and model each of these unidimensional functions as a unidimensional GP with its own mean function and kernel structure~\citep{Duvenaud2013}. We use such a decomposition to model $\alpha(t,r_{it},\ell_{it})$ in the GPPM. Modeling a multi-input function additively, dimension-by-dimension, allows us to more easily understand patterns along a given dimension, and facilitates visualization, as the sub-functions are functions on $\reals$. Furthermore, just as the sum of Gaussian distributions is Gaussian, the sum of Gaussian processes is also a Gaussian process, with a mean function equal to the sum of the mean functions of the component GPs, and its kernel equal to the sum of the constituent kernels. This is called the additivity property of GPs. 

In fact, the additivity property of GPs allows us to do much more than see the connection between these multidimensional and unidimensional function spaces; it also allows us to define a rich structure along a given dimension of the input space. Let us return for the moment to the unidimensional case, with an input $\tau \in \reals$ and a function $f$. By the additivity of GPs, we can model the latent function $f$ as a sum of subfunctions on the same input space,
\begin{equation} f(\tau) = f_1(\tau)+f_2(\tau)+\ldots+f_J(\tau), \end{equation}
and allow each of these subfunctions to have its own mean function, $m_j(\tau)$, and kernel, $k_j(\tau,\tau')$. The mean function and kernel of the function $f$ are then given by $m(\tau) = \sum_{j=1}^J m_j(\tau)$ and $k(\tau,\tau') = \sum_{j=1}^J k_j(\tau,\tau')$, respectively. This allows us to represent complex structure on even a single dimensional input, even when using simple kernels like the squared exponential. We can, for example, allow the different subfunctions to have different squared exponential kernels that capture variability along different length-scales, and add a strictly cyclic kernel to isolate predictable cyclic variability of a given cycle length. It is through this additive mechanism that we represent long-run, short-run, and cyclic trends in the calendar time component of the GPPM.

\subsection{GPPM Dynamics Specification}

The flexibility GP priors provide in nonparametrically representing detailed underlying structures in a function space makes a GP framework ideal for estimating our latent, time-varying function, $\alpha(t,r_{it},\ell_{it})$. Recall that the basic form of the GPPM is:
\begin{equation} \mbox{Pr}(y_{it}=1) = \mathrm{logit}^{-1}\left[\alpha(t,r_{it},\ell_{it})+\x_{it}^{\prime}\bt{\beta}+\z_{i}^{\prime}\bt{\gamma} + \delta_i\right].\end{equation}
We can use the additivity property of GP priors to expand the single latent function into three subfunctions along the three dimensions of interest,
\begin{equation}
\alpha(t,r_{it},\ell_{it}) = \alpha_{T}(t) + \alpha_{R}(r_{it})+\alpha_\mathcal{L}(\ell_{it}),
\end{equation}
and model each of these functions using a separate GP prior. 

In modeling the calendar time dynamics, we again exploit the additivity of GPs. We decompose the calendar time component $\alpha_{T}(t)$ into three sub-functions that can capture long-term trends, short-term variability, and predictable cyclic patterns, such that 
\begin{equation}
\alpha_{T}(t) = \alpha_{T}^L(t) + \alpha_{T}^S(t) + \alpha_{T}^W(t).
\end{equation}
In the above, $L$ denotes the long-run trend component, $S$ denotes the short-run component that captures the impact of transient events, and $W$ denotes the day-of-week effect. We then use unidimensional GPs with kernels that capture these three types of dynamics. Note that, although in this application, day of the week is the only applicable cyclic quantity of interest, patterns of other periodicities can be readily incorporated in our framework. Hence, the final form of the dynamics function is:
\begin{equation}
\alpha(t,r_{it},\ell_{it}) = \alpha_{T}^L(t) + \alpha_{T}^S(t) + \alpha_{T}^W(t) + \alpha_{R}(r_{it})+\alpha_\mathcal{L}(\ell_{it})
\end{equation}
where each of these components is given a unidimensional GP prior. 

For each component except $\alpha_{T}^W(t)$, we use a standard squared exponential kernel. For $\alpha_{T}^W(t)$, we use the strictly cyclic variant of the squared exponential kernel. As the hyperparameters of the different kernels are estimated from the data, the best length-scales that describe the variation in the data are automatically inferred. We differentiate between long-run and short-run calendar components through the value of the length-scale parameter, with a higher length-scale implying correlations in outputs over longer spans of inputs. To prevent label-switching when estimating these latent functions, we constrain the length-scale of the long-run component to be greater than the length-scale of the short-run component.

\subsubsection{Mean Functions}
We use mean functions to weakly encode monotonicity assumptions, and to facilitate forecasting. Specifically, we use a constant mean function for the calendar time components as we have no reason to assume any specific pattern. For the interpurchase time and customer lifetime components, however, past research as well as intuition suggests a decreasing pattern over time: the longer a customer has gone without spending, or the longer since a customer's first spend, the less likely it is for the customer to spend at any given time. We encode this assumption as a mean function of the form $m(\tau)=-\lambda_1(\tau-1)^{\lambda_2},~\lambda_1,~\lambda_2>0$. Note again that this does not \textit{impose} monotonicity, but rather sets a prior for monotonicity. Thus, a randomly drawn function may not exhibit monotonicity.  Further, note that this mean function approaches the constant mean function as $\lambda_1 \rightarrow 0$. Thus, if the data imply a non-decreasing shape, we will do no worse than a flat mean function. As we discussed earlier, the mean function has little effect in areas where the data are dense and therefore its impact in the calibration sample is minimal. However, the mean function can drive out-of-sample forecasts, particularly over a long range. For holdout data that is in close range of the calibration data, forecasts are governed by both the mean function and the length-scale of the covariance kernel. 

\subsection{Full Model}
Apart from the time varying effects, we use $\x_{it}$ to control for the purchase number for customer $i$ at time $t$. In our application, we use a set of dummies to specify these time varying effects. We do not explicitly add other time varying effects because we anticipate that these will be nonparametrically captured by the calendar-time components of the GP. We rely on the nonparameteric specification to automatically represent any unknown effects which may result from information asymmetries or unobserved shocks. We also include time invariant controls for managerially relevant observed heterogeneity, including fixed effects for the acquisition channel of the customer, and random effects across days for when the customer installed the game and when the customer first spent in the game.

In sum, our model is driven by a time-varying additive function describing the spending rate, which has component sub-functions representing predictable strictly cyclic patterns, and long and short-run variability, all of which are given Gaussian Process priors. Together, this amounts to:
\begin{gather}
 \mbox{Pr}(y_{it}=1) = \mathrm{logit}^{-1}\left[\alpha_{T}^L(t) + \alpha_{T}^S(t) + \alpha_{T}^W(t)+\alpha_{R}(r_{it})+\alpha_{\mathcal{L}}(\ell_{it}) + \x_{it}^{\prime}\bt{\beta}+\z_{i}^{\prime}\bt{\gamma} + \delta_i\right], \nonumber  \\
\alpha_{T}^L(t) \sim \mathcal{GP}(\mu,K_{\textrm{SE}}(t,t';\eta_L,\rho_L)), \nonumber  \\
\alpha_{T}^S(t) \sim \mathcal{GP}(0,K_{\textrm{SE}}(t,t';\eta_S,\rho_S)), \nonumber  \\
\alpha_{T}^W(t) \sim \mathcal{GP}(0,K_{\textrm{C}}(t,t^{\prime};\eta_W,\rho_W)), \\
\alpha_{R}(r) \sim \mathcal{GP}(-\lambda_1^R(r-1)^{\lambda_2^R},K_{\textrm{SE}}(r,r^{\prime};\eta_R,\rho_R)), \nonumber \\
\alpha_{\mathcal{L}}(\ell) \sim \mathcal{GP}(-\lambda_1^\mathcal{L}(\ell-1)^{\lambda_2^\mathcal{L}},K_{\textrm{SE}}(\ell,\ell^{\prime};\eta_\mathcal{L},\rho_\mathcal{L})),  \nonumber \\
\delta_i \sim \mathcal{N}(0,\sigma_{\delta}^2), \nonumber 
\end{gather} 
subject to the constraints: $\rho_S<\rho_L$ (which without loss of generality differentiates between short- and long-run calendar effects), and $\lambda_1^R,\lambda_1^\mathcal{L},\lambda_2^R,\lambda_2^\mathcal{L}>0$. 

We need to impose identification restrictions because of the additive structure of our model. Sums of two latent functions, such as $\alpha_1(t)+\alpha_2(t)$, are indistinguishable from  $\alpha_1^*(t)+\alpha_2^*(t)$, where $\alpha_1^*(t)=\alpha_1(t)+c$, and $\alpha_2^*(t)=\alpha_2(t)-c$ for some $c \in \reals$, as both sums imply the same purchase probabilities. To address this indeterminacy, we set the initial function value (corresponding to input $\tau=1$) to zero for all of the latent functions, except for $\alpha_{T}^L(t)$. In this sense, $\alpha_{T}^L(t)$ captures the base spending rate for new customers, and the other components capture deviations from that as time progresses. We also impose the usual dummy coding restrictions on the fixed effects and a zero mean for random effects.

\subsection{Benefits of the GP Approach}

As GPs are novel to the marketing literature, it is worthwhile to explore briefly the rationale for using them in modeling latent functions. The nonparametric approach offers many benefits in that it makes minimal assumptions about the forms of the latent functions, it flexibly accommodates situations with differing spending patterns, and  requires no manual adjustment by a marketing analyst to handle different product contexts. GPs offer many other benefits over other nonparametric models. They allow for a structured decomposition of a single process, such as the the calendar time process, into several subprocesses via the additive property. This additive formulation facilitates a rich representation of a dynamic process via a series of kernels that can capture patterns of different forms (e.g., cyclic vs. non-cyclic) and operate at different time scales. Yet, as the sum of GPs is a GP, the specification remains identified, with a particular mean and covariance kernel. Achieving a similar representation with other nonparametric methods that are based on fixed effects or splines is either infeasible or more difficult. Finally, the fundamental link between GPs and the multivariate normal distribution makes estimation straightforward, especially from a Bayesian perspective.

\section{Empirical Application}

\subsection{Data}
The data consist of spend incidence logs for two games from one of the world's largest video game companies.\footnote{There is no personally identifiable information in our data; player information is hashed such that none of the data we use or the results we report can be traced back to the actual individuals.  We also mask the identification of the company per their request.} The games are free-to-play mobile games, an expanding and hugely profitable segment of the video game industry. In free-to-play settings, users can install and play video games on their mobile devices for free, and are offered opportunities to purchase within the game. These spend opportunities typically involve purchasing in-game currency, like coins, that may subsequently be used to progress more quickly through a game, obtain rare or limited edition items to use with their in-game characters, or to otherwise gain a competitive edge over non-paying players. Clearly, the nature of these purchases will depend on the game, which is why it is important for a model of spending behavior to be fully flexible. We cannot name the games here because of non-disclosure agreements. Instead, we use the general descriptors Life Simulator (LS) and City Builder (CB) to describe the games. 

The games and ranges of data used were selected by our partner company, in an effort to understand spend dynamics over specific periods of time. We use a random sample of 10,000 users for each of the two games. Each sample is drawn from users who installed the game within the first 30 days, and spent at least once during the training window. 

In the Life Simulator (LS) game, players create an avatar, then live a digital life as that avatar. Purchases in this context can be rare or limited edition items to decorate or improve their avatar or its surroundings. Often times, limited edition items are themed according to holidays such as Christmas or Halloween. Our data come from a 100 day span of time covering the 2014 Christmas and New Year season. In the City Builder (CB) game, players can create (or destroy) a city as they see fit. Customers make purchases to either speed up the building process or to build unique (or limited edition) additions to their cities. Our data come from an 80 day period of time at the start of 2015, at the tail end of the Christmas and New Year holidays. 

\begin{figure}[p!]
\centering
\makebox[\textwidth][c]{\includegraphics[scale=0.75]{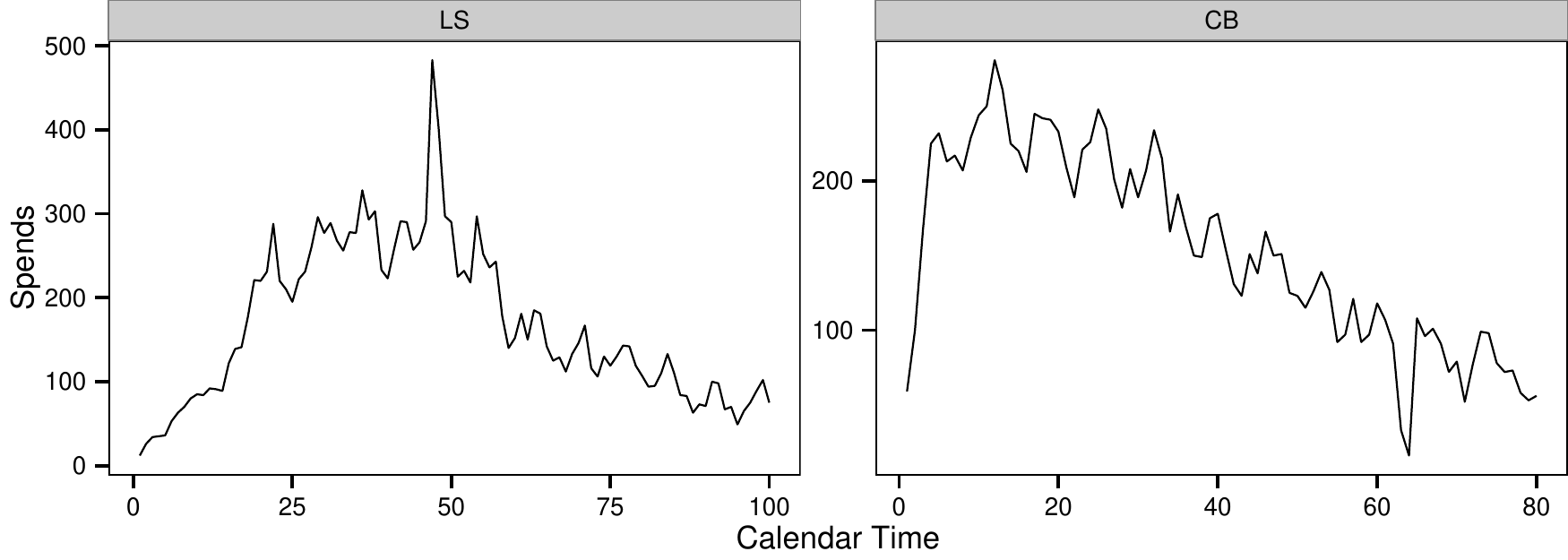}}
\caption{The spends-by-day time series.}
\label{fig:sbd}
\end{figure}

The time series of spending for the two games are shown in Figure~\ref{fig:sbd}. From these figures, it is difficult to parse out what exactly is driving the aggregate pattern of purchases.  The figure includes customers who installed the game any time within the first thirty day window. Typically, customers are most active when they start playing a game, so we expect to see more spending  in the first 30-40 days simply because there are likely more people playing in that period, and new players are entering the pool of possible spenders. It is unclear what else underlies the aggregate spends. We see many peaks and valleys in spending over the entire time horizon and without deeper analysis, it is difficult to discern which  ``bumps" in the plots are meaningful, and which represent random noise. For example, if 5,000 players are active at any given day, then a jump of 50 spends in may represent a random fluctuation. In contrast, if only 1,000 players are active, the same jump of 50 spends may be very meaningful.  In other words, the significance of a particular increase in spending depends on how many customers are still actively spending at that time, and on other patterns that may operate at the individual-level. Thus, it is important to develop a model-based understanding of the underlying dynamics.

\subsection{Model Estimation}

We estimated the model using 8,000 of the 10,000 customers from our sample. The rest form a holdout sample to assess the generalizability of our model. We employ a fully Bayesian approach for estimation and use the NUTS~\citep{Hoffman2011} variant of Hamiltonian Monte Carlo (HMC) method~\citep{Neal2011} as implemented in the Stan probabilistic programming language (Stan Development Team, 2015). NUTS allows for automatic adaptation of the step-size parameter and the number of leapfrog steps of the HMC algorithm. HMC in general, and NUTS in particular, avoid the random walk behavior of ordinary Metropolis-Hastings procedures and therefore explores the posterior in a very efficient fashion. We assess convergence through the $\hat{R}$ statistic~\citep{Gelman1992}. Even though our model converges in as few as 2,000 HMC iterations, we use results based on 4,000 HMC draws. 

\subsection{Results and Managerial Insights}

The GPPM offers a visual and highly general system for customer base analytics that is driven by nonparametric latent spend propensity functions. These latent curves form the primary output of the model, and their posterior median estimates are displayed in Figure~\ref{fig:ls_dashboard} for LS, and in Figure~\ref{fig:cb_dashboard} for CB. We call these figures the GPPM ``dashboards," as they visually represent latent spend dynamics, and could serve as a highly effective basis for an automated decision support system. These latent curves, combined with our estimated fixed and random effects, result in an excellent fit to the data, both in and out-of-sample. 

In the subsequent sections, we first showcase the validity of the model through its fit to the data, and illustrate how the dynamic components of the model influence that fit. We then discuss the managerial insights that can be gleaned from the dashboard for both understanding a given game, and comparing across games with distinct customer bases. Finally, we discuss the hyperparameters that drive these latent curves, and the estimated fixed and random effects that complement the dashboard's explanatory power.

\subsubsection{Model Fit}

\begin{figure}[p!]
\centering
\makebox[\textwidth][c]{\includegraphics[scale=0.75]{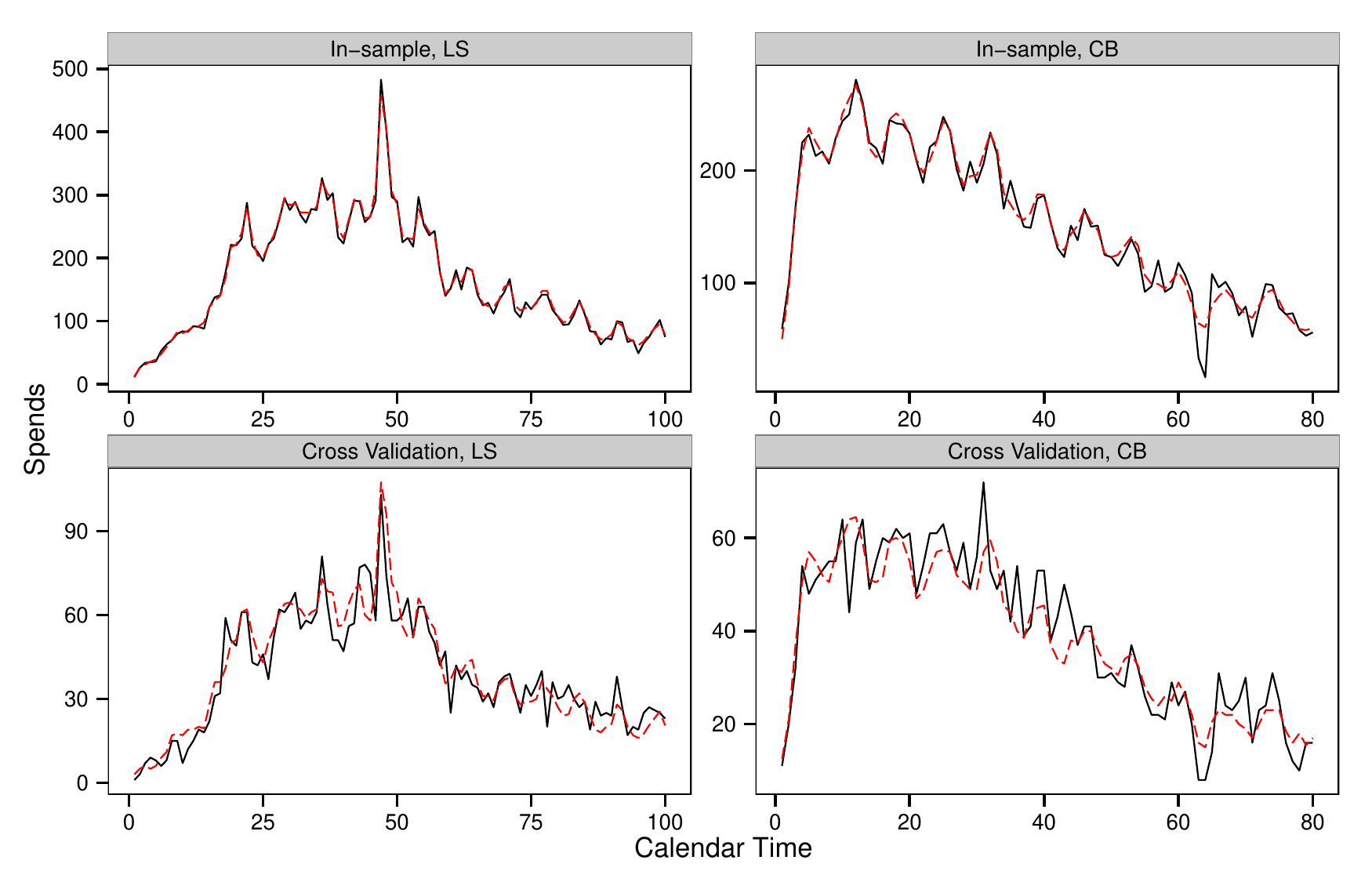}}
\caption{Posterior simulated fits. We used the estimated parameters across many posterior draws of the HMC sampler to simulate spending, and plotted it against the true spends by day. In the top row, we do this with the original data. In the bottom row, we do this on the cross validation data of 2,000 spenders.}
\label{fig:fit_insample}
\end{figure}

First, we look at model fit, both in the calibration sample of 8,000 customers and in the holdout sample of 2,000 customers. A closed form expression is not available for the expected number of aggregate counts in the GPPM. We therefore  simulate spending from the posterior predictive distribution by utilizing the post convergence HMC draws for the latent curves and the fixed and random effects. The top row of Figure~\ref{fig:fit_insample} shows the actual spending and the median simulated purchase counts (dashed line) for the two games. We see that the fit is exceptional, and tracks the actual purchases almost perfectly, especially for LS. This is not surprising: because we capture short-run deviations in the probability of spending, we essentially capture the residuals from the smoother model components in a probabilistic sense. That is, the short-run component captures any probability that is ``left-over" from the rest of the model. 

To test that the model does not overfit the in-sample day-to-day variability, we  explore the simulated fit in the validation sample. The bottom row of Figure~\ref{fig:fit_insample} shows that the fit to this sample is still excellent, although not as perfect as in the top row. While the probabilistic residuals from the calibration data are not relevant for the new sample, much of the signal present in the calendar time effects, trends, and the recency and lifetime effects continue to matter, thus contributing to the fit. 

\paragraph*{Fit Decomposition}

To better understand how the latent curves in the dashboard contribute to the fits seen in Figure~\ref{fig:fit_insample}, we now break down that fit along the latent dimensions of calendar time, recency, and lifetime. For brevity, we focus on the near perfect fit of LS in the top-left panel of Figure~\ref{fig:fit_insample} and examine how the fit changes when different components of the model are muted, one at a time. We ``mute'' a component by setting  it to its mean over time (which for many components, is zero by design). Note that we do not re-estimate a model when we mute a component. Instead, muting allows us to see how much of the overall fit is driven by a given component. 

\begin{figure}[p!]
\centering
\makebox[\textwidth][c]{\includegraphics[scale=0.75]{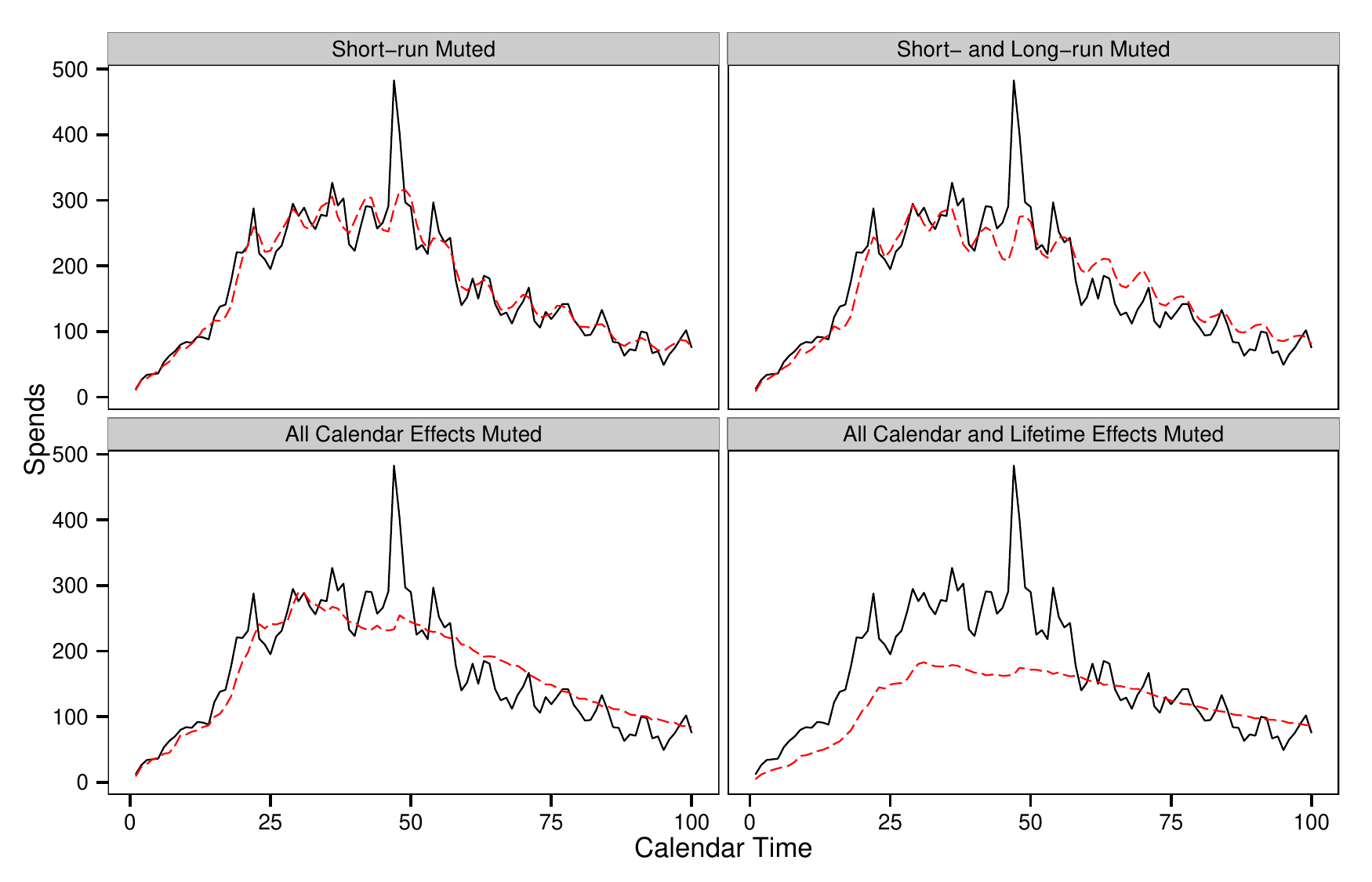} }
\caption{Life Simulator posterior simulated fit, with various components muted. By muted, we mean one of the components of the latent $\alpha(t)$ function set to its estimated mean value, and spending simulated, effectively removing the variance in simulated fit that can be attributed to that component of the GPPM.}
\label{fig:ls_fit_decomp}
\end{figure}

The top-left panel of Figure~\ref{fig:ls_fit_decomp} shows the results when the short-run component is muted. It is immediately obvious from the figure that the short-run component serves to capture the deviations from the smoother long-run, recency, and lifetime effects. It therefore provides event detection---deviations from what is expected are reflected as spikes in the short-run curve. It is also obvious from the very good fit of the muted short-run model that much of the variance in day-to-day spending is still captured even when the short-run component is shut off. This again just confirms what we saw in the cross-validation fit, and what we claimed before: while many of the tiny deviations in the short-run curve simply represent noise, these do not drive model fit. Instead, the influence of the short-run process on the fit is driven by the large short-run deviations, which as we will see are typically linked to events of interest.

Unlike the short-run calendar time effects that capture very little of the variance in daily spending outside of sudden big peaks or dips, the long-run calendar time dynamics explain a large extent of the variation. The top-right panel in Figure~\ref{fig:ls_fit_decomp} shows the results when short and long-run effects are muted. The loss of fit here can be better understood by looking closely at the dashboard in Figure~\ref{fig:ls_dashboard}.  We notice a distinct shift in the long-run component of the dashboard around time 50: there was a period of heightened spending that then fell to a relatively constant but lower base level. As we will see, this dip corresponds to a post-Christmas slump. Thus, if calendar time dynamics are muted, (i.e, assumed to be a constant), then we underestimate spending in periods when the true calendar time effect is high (e.g, in the beginning of the data, and specially when the calendar time components peak before time 50), and overestimate spending when it is low. Returning to the top-right panel of Figure~\ref{fig:ls_fit_decomp}, that is exactly what we find.

Most previous models of customer base analysis have almost exclusively focused on the non-calendar time aspects of spending: recency and lifetime. If we fully mute all of the calendar time components of the model, i.e., the cyclic day of the week effect as well as the short and long-run components, we approximate what we refer to elsewhere in the paper as the reduced GPPM (or rGPPM). The rGPPM is a direct analogue of these recency and lifetime models. The effect of muting all calendar time effects is shown in the bottom-left panel of Figure~\ref{fig:ls_fit_decomp}. We see that accounting only for recency and lifetime dynamics yields a reasonable estimate of the mean level of spending over time. However, when we compare this fit to that in the top row, where the predictable cyclic component is still part of the model, we realize that a large part of what would be considered noise under recency and lifetime models is actually systematic variation. Specifically, the top two panels of Figure~\ref{fig:ls_fit_decomp} show that many of the hills and valleys of day-to-day spending can be predicted precisely by isolating a strict cyclic calendar time pattern. In other words, the actual spending rate is not as noisy as it appears to be in models that focus only on recency and lifetime.

Finally, we consider how much the lifetime component contributes. Previous research suggests this to be a significant driver and indeed our analysis supports this view. The bottom-right panel of Figure~\ref{fig:ls_fit_decomp} shows the fit with the lifetime effect muted, along with all of the calendar time effects. We see that while the overall shape is still loosely captured, all of the nuance is lost, and the fit is significantly poorer than with the lifetime component included. This is not surprising given the literature on customer base analysis that emphasizes the importance of latent churn processes, of which our lifetime effect is a direct analogue.

\subsubsection{Dashboard Insights}

\begin{figure}[p!]
\centering
\makebox[\textwidth][c]{\includegraphics[scale=0.75]{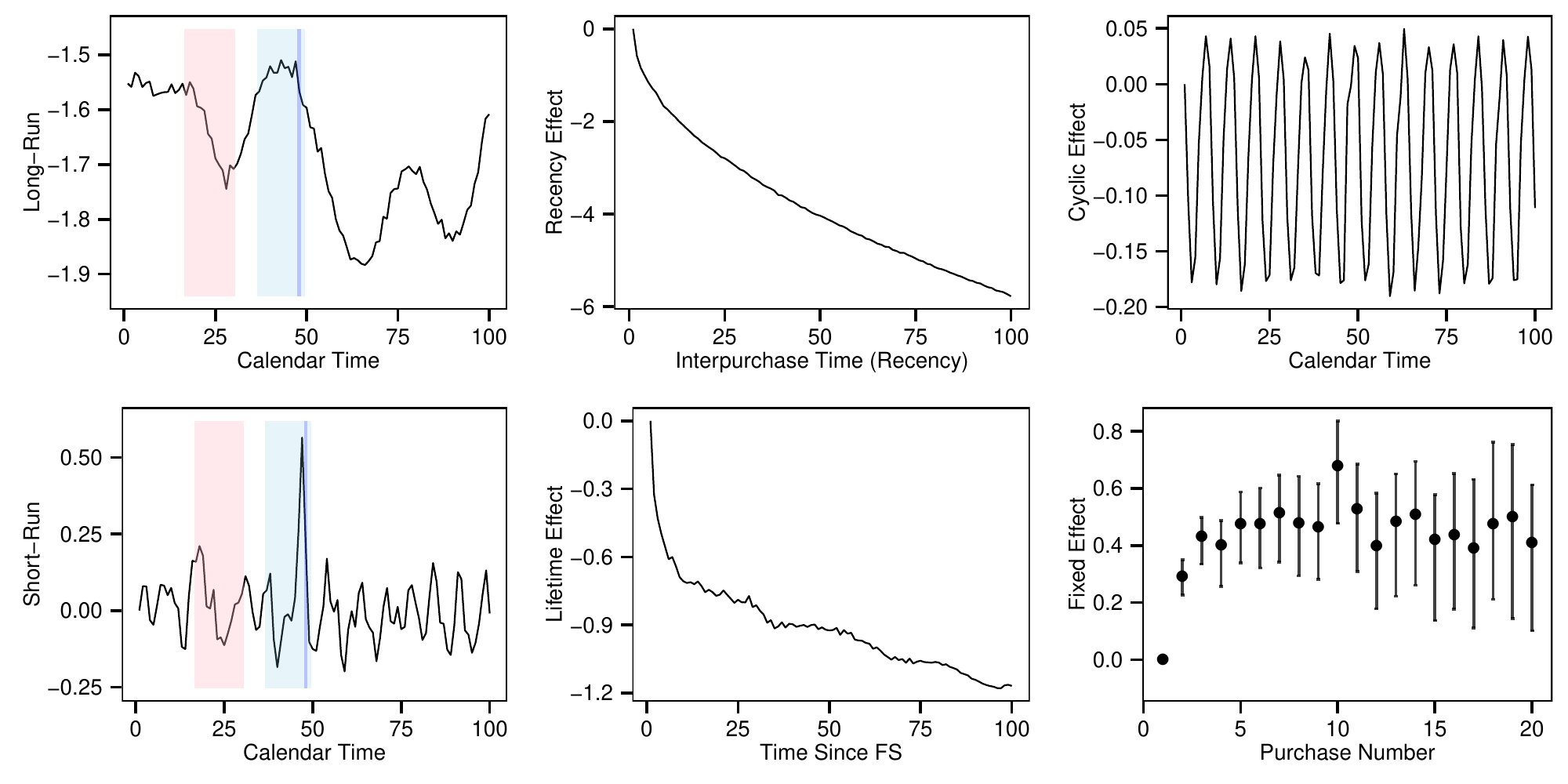}}
\caption{Life Simulator Dashboard. Highlighted areas are ``events of interest" from the company. From left to right, we have a charity promotion, where the game turned red and all proceeds benefited a charity, and the holiday season, where a Christmas holiday-themed quest was added to the game and where Christmas day is the darker area just before day 50.}
\label{fig:ls_dashboard}
\end{figure}

\begin{figure}[p!]
\centering
\makebox[\textwidth][c]{\includegraphics[scale=0.75]{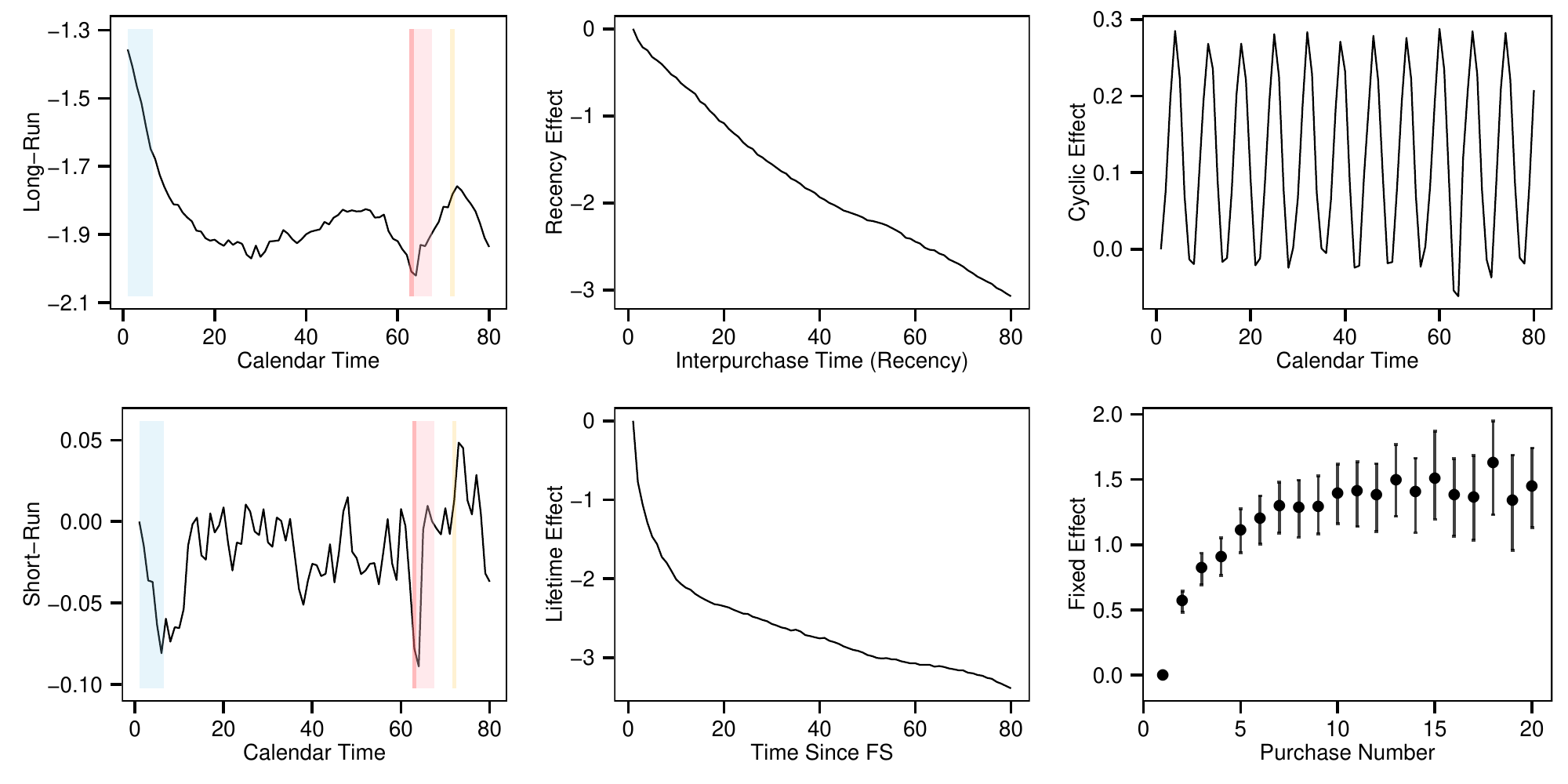}}
\caption{City Builder Dashboard. Highlighted areas are ``events of interest" from the company. From left to right, we have the tail end of the holiday season, an update intended to promote repeat spending (where the update itself is darkly marked, and a general range of time where an impact is expected is lightly marked), and a error in the app store.}
\label{fig:cb_dashboard}
\end{figure}

While fit validates the utility of the GPPM, the key motivation of the model is to provide managers with a model-based decision support system that visually captures effects of interest. Thus, the key output of our model is the GPPM dashboard (Figures~\ref{fig:ls_dashboard} and~\ref{fig:cb_dashboard}), which portrays the median posterior estimates of the latent propensity functions. These latent spend propensity curves are readily interpretable, even by managers with minimal statistical training. We illustrate here the insights managers can obtain from these model-based visualizations. 

\paragraph{Calendar Time Effects}

Events that happen in calendar time are often of great importance for managers, but their impact is often omitted from typical customer base models. The GPPM includes these effects nonparametrically, and the additive decomposition of the calendar time effects into multiple subcomponents allows an intuitive representation of the impact of events. The effects take the form of easily assessed curves, and as the model is specified on a  probability scale, the curves are comparable across games. We use the data from our two games to illustrate what a manager can lean from the GPPM dashboards.

\textit{LS Dynamics}~~~The shaded areas in the calendar time components of the dashboard in Figure~\ref{fig:ls_dashboard} correspond to events that are of interest to the company. Two events of note  occurred in the span of the data. The first highlighted area from $t=17-30$ corresponds to a period in which the company made a game update, introduced a new game theme involving a color change, and also donated  all proceeds from the purchases to a charitable organization. The second shaded area, around $t=37-49$, corresponds to another game update that added a Christmas-themed quest to the game, with Christmas itself falling on the darker highlighted day $t=48$, right before the end of the holiday quest. We can use the GPPM to evaluate whether any notable change in spend propensities happened during these times. It is crucial to note that none of these event periods were specifically incorporated in the GPPM; instead, we can use the GPPM to evaluate their effects by examining the dashboard in these regions. We can also use the dashboard to flag events that disrupt spending that may be outside the awareness of managers.

First, consider the charity update period: the long-run curve shows that the spending rate was relatively high at the beginning of the period, but declined consistently throughout. Thus the update appears to be negatively correlated with repeat spending. The short-run curve perhaps offers a better story: there is a small bump in spending corresponding to the start of the new update, but relative to the noise elsewhere in the plot, this does not appear to be very meaningful. 

Next, we investigate whether the Christmas update had a more substantial effect. Indeed, the long-run curve shows that spending was elevated and rising during the time of the update. However, we also notice that this rise in spending started before the range of the update, and in fact is it quite possible that  the general level of spending could just be returning to its level before the charity update. Hence, the effect of the actual Christmas update is ambiguous. Overall, it appears that the holidays do have an impact on spending, given the quite distinct bump in long-run spending throughout the month of December, which began at $t=24$, and the significant dropoff in spending that started almost immediately following the holiday season. The short-run curve shows a distinct and interesting impact of Christmas Eve: while Christmas appears to be relatively uneventful, there is a large and unambiguous spike in spending on Christmas Eve. The fact that the GPPM can uncover such a Christmas Eve effect showcases its ability  to account for unanticipated effects that one cannot a priori accommodate via covariates. As we elaborate in Section 4 of the paper, a spike such as this results in loss of fit and diminished predictive ability for standard probability models that do not incorporate calendar time effects. 

\textit{CB Dynamics}~~~ The shaded areas of the CB dashboard in Figure~\ref{fig:cb_dashboard} again correspond to events of managerial interest, as provided by our partner company. The start of the data window ($t=1-6$) coincides with the tail end of the holiday season, from December 30 to January 4. Another event begins at $t=63$, when the company launched a permanent update to the game to encourage repeat spending. We highlight five additional days after that update to signify a time period over which significant post-update activity may occur. Finally, at $t=72$, there was a crash in the app store. 

We see, as in the previous game, that the spending level in the holidays ($t=1-6$) was quite high and fell dramatically subsequently. Spending over the rest of the year was relatively stable. The update that was intended to promote repeat spending  had an interesting effect: there was an initial drop in spending, most likely caused by reduced playtime on that day because of the need for players to update their game or because of an error in the initial launch of the update. After the update, an uptick in long-run spending is observable, but this was relatively short-lived.\footnote{We note that statistically, there appears to be a bit of lack of separation between the short and long-run components in this case, as there is a rather noticeable dip in spending corresponding exactly to the dip in spending in the short-run curve. This arises because we are attempting to sort all of the effects along two characteristic length-scales for the sake of interpretability.} Finally, we find no effect for the supposed app store crash, which in theory should have prevented players from purchasing for the duration of the crash. It is plausible that the crash was for a short duration or occurred at a time when players were not playing. Again, we make these inferences post hoc, knowing the time periods of interest to the company, but without using the events as input to the model.

\textit{Cyclic Effects and Cross Game Comparisons}~~~We find that the day of the week (DOW) cyclic effect is less powerful in LS than in CB. In fact, this is even evident in the aggregate time series themselves (Figure~\ref{fig:sbd}). As we will see later, this makes the spending more predictable in CB, and therefore aids the forecasts of future spending.

How do the other calendar time effects in LS compare to those in CB? As the GPPM model relies on a probabilistic specification of the spending propensity, the results from different dashboards are directly comparable. Our analysis shows that both games experienced a higher level of spending during the holiday season. We notice that the drop in spending in CB after the holidays was greater than that in LS. Across both games, we find that the effects of updates appears to be unpredictable: while in CB there was no appreciable uptick in spending for the ``repeat spender" update,  there was a somewhat appreciable uptick for the charity update in LS, followed by a substantial downturn. As many of these effects and especially their shapes are difficult to predict a-priori, this showcases the benefits of a nonparametric approach.

\paragraph{Event Detection} Managers are often interested in real-time detection of events of interest. The short-run function is capable of capturing deviations in short-run spending due to an event of interest. That is, if something disrupts spending for a day, such as a crash in the payment processing system, or an in-game event, it will be reflected as a spike in the short-run function, as evident for example in the Christmas Eve effect in LS above. It is natural to ask how long it takes for the model to detect such a deviation in spending.

\begin{figure}[p!]
\centering
\makebox[\textwidth][c]{\includegraphics[scale=0.75]{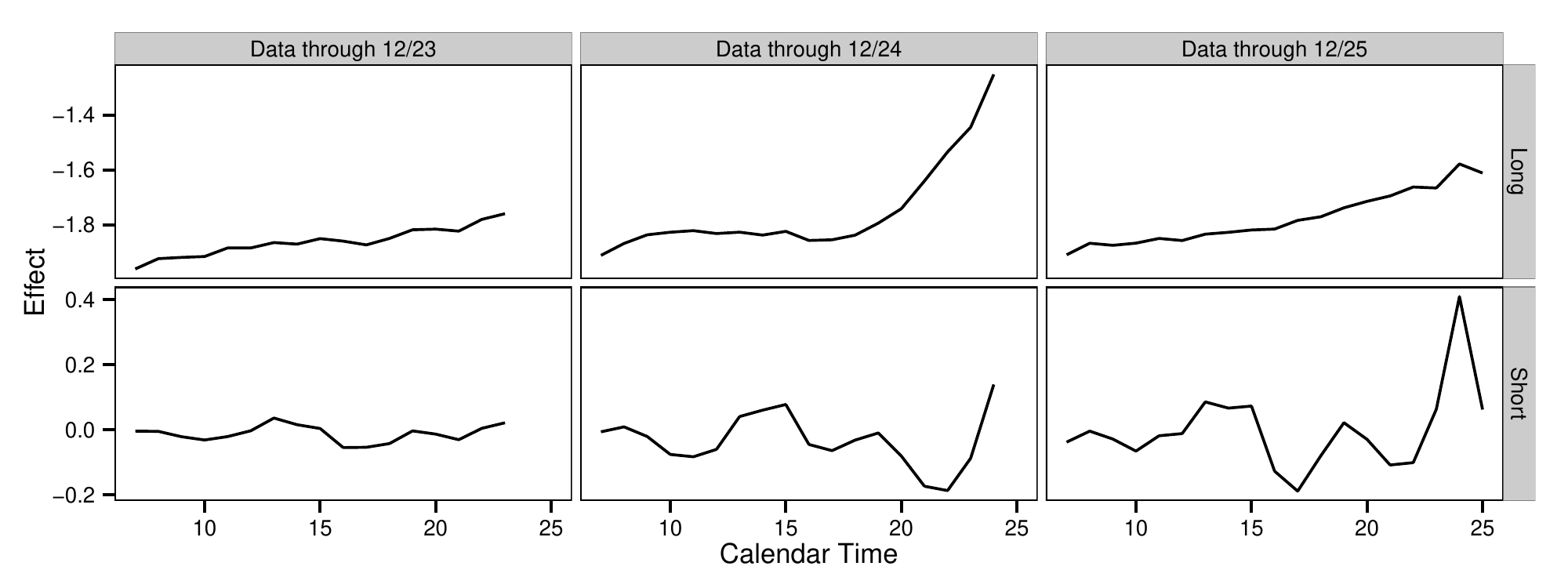}}
\caption{Event detection. The GPPM separates out the short and long-term effects of the holiday season, indicating a short-run disturbance for Christmas Eve, detectable after only 2 days. }
\label{fig:detect}
\end{figure}

The GPPM estimation process works by decomposing a given process along sub-functions with differing length-scales. As such, when there is a deviation, it must learn the relevant time scale for the  deviation---here, either short or long-term---and then adjust accordingly. We illustrate this dynamically unfolding adjustment process in Figure~\ref{fig:detect} by estimating the model using progressively more data over the range 12/23/2014 to 12/25/2014. The different columns of the figure show how the long-run (top row) and the short-run (bottom row) components vary when data from each successive day is integrated into the analysis. To clearly illustrate the adjustment process, we set the abscissa for each subplot in the figure to reflect the actual date, i.e, the specific day in December. The second column shows the impact of adding the data from  Christmas Eve. An uptick in spending is apparent, but the GPPM cannot yet detect whether this uptick will last longer or just fade away. The day after (third column), it becomes clear from looking at the long-run and short-run plots that the effect was only transient.

The above example shows that  the GPPM not only automatically captures the ``correct" form of the Christmas Eve effect but it also makes this effect immediately apparent via the updated dashboard. This capability can be immensely valuable to managers in multiproduct firms where information asymmetries abound. For example, in  digital contexts, product changes can sometimes be rolled out without the knowledge of the marketing team. Similarly, disruptions in the distribution chain can occur with little information filtering back to marketing managers. The GPPM can capture the impact of such events automatically and quickly.

\paragraph{Recency and Lifetime Effects} While the calendar time components are the real novelty of the GPPM, the recency and lifetime effects are nonetheless important. In this case, the recency and lifetime effects are very smooth and decreasing as expected. For managers, this simply means that the longer someone goes without spending, and the longer someone has been a customer in these games, the less likely that person is to spend. The recency effect is consistent with earlier findings and simply indicates that if a customer has not spent in a while, he or she is probably no longer a customer. The lifetime effect is also expected, especially in the present context, as these games offer a limited number of things to purchase, and customers are more likely to branch out to other games with the passage of time. 

\begin{figure}[p!]
\centering
\makebox[\textwidth][c]{\includegraphics[scale=0.75]{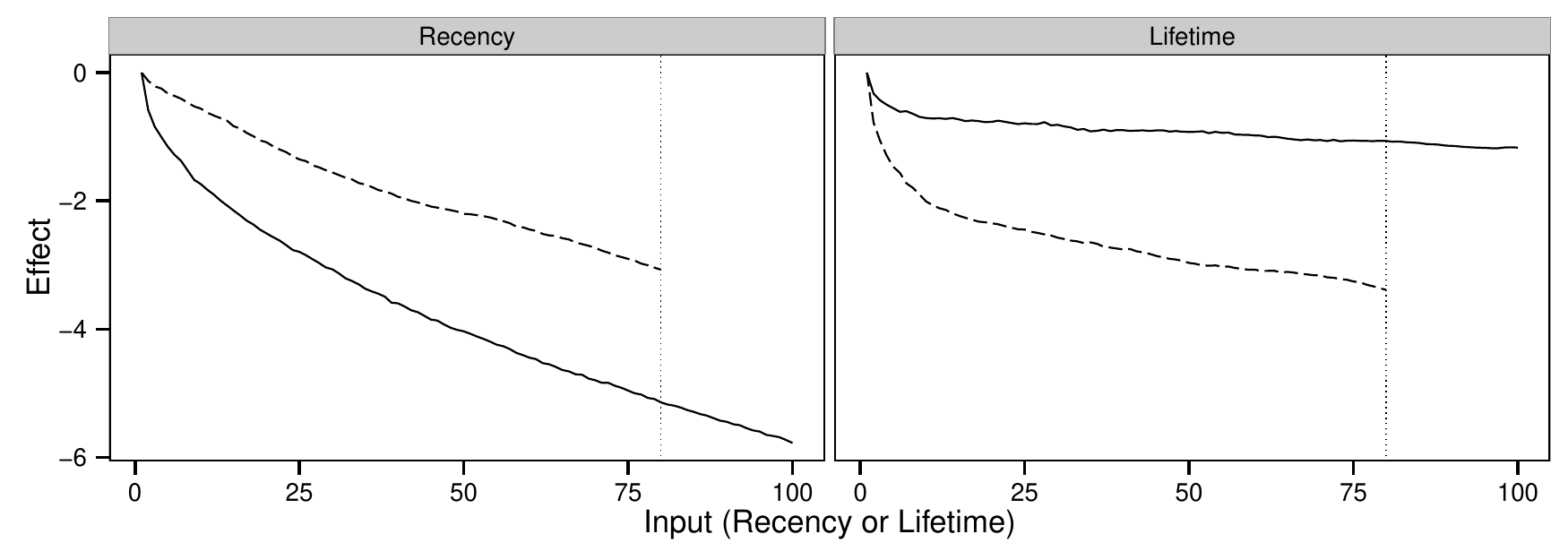} }
\caption{Comparison of the lifetime and recency effects across the games. Here, the curves from LS are solid, and the curves from CB are dashed. Their superposition shows the effects are of varying natures and importances across the games.}
\label{fig:rl_comp}
\end{figure}

More interesting are the rates at which these decays occur, and how they vary across the games. These processes appear to be fundamentally different in the two games as can be seen in Figure~\ref{fig:rl_comp}.  In LS, the recency effect has a large impact, whereas the lifetime effect assumes a minimal role. In contrast, in CB, the lifetime effect is large, while the recency effect is less important, relative to LS. The shapes of these curves also have implications for player behavior. In LS, the recency effects hits hard, meaning that if a player does not spend soon after the first purchase, then he or she is unlikely to spend again. In contrast, in CB, the recency effect decays gradually, almost linearly, whereas the lifetime effect implies that players appear to quickly tire of spending. This could be because there is less to buy in CB, or because the returns to spending are less in this game than in LS.

\paragraph{Purchase Number} The dashboards also show the time-varying effect of the  current (repeat) spend number of the customer. We find here a very distinct effect, which is more obvious for CB than for LS: the probability of purchasing again is much higher for people at higher purchase numbers. This likely reflects additional customer heterogeneity, in that a customer who spends five times is different from someone who spends one time. This effect is highly concave in both games, a pattern that recurs even in other applications with our partner company. It also speaks to the idea that the repurchase curves and effects may differ across different levels of spending.

\paragraph{Integrated Customer Insights} While each of the components of the model provides managerially relevant insights in isolation, they are also nested within a single spend propensity, and as such are more interesting when considered jointly. The overall framework of the GPPM is a discrete time hazard model, where the latent propensities are linked to spend probabilities through an inverse logit link. The input space of the inverse logit link is the sum of the output spaces of the individual GPPM components, meaning the different latent processes and effects additively compete in an effort to explain the purchase propensities in the data. As the purchase propensity function is nonlinear, the effect of a given model component cannot be understood in isolation. All components interact in an implicit sense, as the impact of an increase in one component on the probability of purchasing implicitly depends on the level of the other components. For instance, if a customer has not purchased for a very long time, and thus has a very low recency effect, then the other model components will only be able to exert a minimal influence on the getting the person to spend. In other words, if a customer is ``dead", the calendar time components cannot have an impact on the customer.  

The notion of input competition stems from the fact that the input to the inverse logit function can meaningfully impact the probability of spend only in a restricted range, i.e, over the interval $[-5,5]$. Thus, if a function component spans a large range of the input space of the inverse logit link, it is an important determinant of the spend rates. We therefore can conclude in the LS game that the recency effect is hugely important, while the lifetime effect and the calendar time effect are less important. Similarly, the purchase number effect and the cyclic component have a weak impact. In contrast, for CB, recency and lifetime effects span a similar range, with the lifetime component appearing a bit more important. The cyclic and purchase number effects are also much more pronounced. As the span of the long-run calendar time dynamics is not much greater than in LS, we can conclude that the recency, lifetime, and purchase number effects are much more important for determining spend dynamics in CB than LS.

In summary, we have seen from the above that the GPPM weaves together the different model components in a probabilistic framework and offers a principled approach for explaining the aggregate purchase patterns based on individual-level data. The visualizations generated by the GPPM are not the result of ad hoc data smoothing, but arise from the structural decomposition of the spend propensity through the different model components. By jointly accounting for the different subprocesses that operate along multiple dimensions and at multiple time scales within a dimension, the GPPM is able to assess the true magnitude of an event and statistically parse out the true nature of an event.

\subsubsection{Parameter Estimates}

While the latent propensity functions are the primary output of the model, these latent functions are estimated through GP priors, which are in turn specified using a set of hyperparameters. In this section, we briefly discuss the estimates and insights derived from these hyperparameters and from the other parameters in the model, including the random and fixed effects.

Table~\ref{tab:hyperpars} shows the hyperparameter estimates for the GP kernels of the different processes. Recall that the amplitude of the squared exponential kernel indicates the variance around the mean function, while the characteristic length-scale signifies the smoothness of the function. We find interesting differences in the parameters across the two games. In LS, the long-run and short-run components have relatively similar amplitudes and as these components have flat (zero) mean functions, the amplitudes represent the extent of the variation along the ordinate. The results therefore imply that the two processes are roughly of equal importance in determining the dynamics. In contrast, the long-run amplitude for CB is much greater than that of the short-run component, implying that long-run dynamics are much more important in CB than in LS. We also see differences in the amplitudes of the recency and lifetime components across the games. Specifically, the amplitudes are very small for LS, but larger for CB. This implies that the CB estimated recency and lifetime functions veer more from their estimated mean functions than the LS functions do. 

Table~\ref{tab:pars} presents the other parameters estimated from the model. The hyperparameters for the  mean function of the GPs are displayed in the first section of the table. While the mean function for the long-run calendar time component is simply a constant $\mu$, we use a functional form $-\lambda_1 (\tau - 1)^{\lambda_2}$  for the recency and lifetime functions to encode our beliefs regarding the likely shape of these components. The second section of the table contains variance estimates for the random effects, including unobserved heterogeneity, first purchase date, and install date. While the random effects for the purchase date and install date do not appear to vary very much across days, the variance of the latent baseline spend propensity $\delta_i$ is estimated to be quite high, indicating that customers do appear to differ substantially in their base spending propensities. This variance reflects a highly skewed distribution of unobserved heterogeneity with a long right tail of valuable heavy spenders.

The third section of Table~\ref{tab:pars} reports the posterior summaries for the fixed effects for the different customer acquisition channels used by the company, which include paid advertising channels, social networks, existing customers of the company who came from other games (denoted WC for ``within company"), and organic arrivals not attributed to a given source. We find for LS that WC customers are the most valuable. Customers drawn from Social 1 and those that arrived organically are also more valuable compared to those who were acquired via Ad 1. This is interesting in that customers acquired for free appear to be more valuable relative to those from paid advertising channels. These conclusions, however, do not carry over to  CB, where we find no significant differences across channels, except for Ad 4, which appears to draw the least valuable customers. The much larger right tail in the unobserved heterogeneity distribution for CB is consistent with the above observation that the observed covariates have limited influence in this game. It is important to note, however, that the width of the posterior interval for an effect also depends on the number of spenders acquired via that channel and constrains our ability to detect differences across channels.

\section{Predictive Ability and Model Comparison}

Apart from having an interest in understanding spending dynamics, managers also value the ability to forecast future purchasing activity. Although the primary strength of the GPPM is in uncovering latent dynamics, it also does very well in predicting future spending. Its forecasting ability is mostly driven by those systematic components that are inherently predictable. These include the recency, lifetime, and cyclic calendar time functions. The long-term and short-term calendar time components, in contrast, capture random shocks or reflect marketing events which are unpredictable a priori. While recency and lifetime are typically incorporated in most customer base models, the inclusion of the cyclic component and the nonparametric characterization of even its predictable components allows the GPPM to significantly outperform benchmark models in predictive ability.

We compare predictive performance of the GPPM with that of a number of benchmark models. We focus on both in-sample and out-of-sample performance and therefore estimate the benchmark models as well as reestimate the GPPM by truncating our original data from 8,000 people along the calendar time dimension. In particular, we set aside the last 30 days of calendar time activity to form a holdout dataset to test predictive validity. Forecasting with the GPPM involves forecasting the latent functions that comprise it. In forecasting these latent functions, we use the predictive mechanisms outlined in Section 2.1.

Predictive performance depends primarily on our ability to forecast the recency and lifetime components for the observations in the holdout data. As the holdout data is constructed by splitting the original dataset along the calendar time dimension, a substantial number of the observations in the holdout data contain lifetime and recency values that are within the observable range of these variables in the calibration dataset. This is true for observations belonging to newly acquired customers. However, for the oldest customers, the recency and lifetime curves need to be forecast. 

\subsection{Benchmark Models}

At its core, the GPPM is a very general discrete hazard model and as such it can be compared to other hazard models for interpurchase times~\citep{Helsen1993,Jain1991,Gupta1991,Seetharaman2003}. Similarly, given its reliance of recency and lifetime dimensions of spending, the GPPM is closely related to traditional customer base analysis models for non-contractual settings of the ``buy-till-you-die'' (BTYD) vein, such as the Pareto-NBD~\citep{Schmittlein1987}, BGNBD~\citep{Fader2010} and their variants. In fact, we argue that the GPPM in some sense generalizes this class of models to include calendar time effects, and different varieties of recency and lifetime effects. Given this nesting and the success of the BTYD models at both explaining and forecasting spending, they are of primary interest for comparison. 

In summary, we use the following benchmark models, representing both of these classes of competing models, along with subcases of the GPPM: 
\begin{description}
	\item[Log-Logistic] The first model we consider is a log-logistic hazard model without time-varying covariates. This an ``automatic" model, similar to the GPPM, in that it does not require any covariate inputs from the analyst. In estimating this model, we use the same set of fixed and random effects  as in the GPPM. We choose the log-logistic hazard as it can flexibly represent both monotonic and non-monotonic hazard functions. 
	\item[Log-Logistic Cov] This generalizes the log-logistic hazard model by using time varying covariates specified by our corporate partner, as described in Section 3. The model uses indicator variables to cover time windows of events of interest. 
	\item[BGNBD] We choose the BGNBD as a representative of the BTYD class of models. In the BGNBD, interpurchase times are modeled by an exponential distribution and customers are assumed to have heterogeneous spending rates. The lifetime effect is captured by the possibility of customer death after each purchase, and customers are assumed to have heterogeneous death probabilities.
	\item[rGPPM] This is a restriction of the GPPM in which all the calendar time components are replaced with a constant. This makes the model directly comparable to the BTYD models where only recency and lifetime effects are modeled.\footnote{We estimated the BGNBD with no effects for observed heterogeneity. Similarly, we estimated the rGPPM both with and without these effects. There is no meaningful difference in the fit statistics across both versions of the rGPPM, as supported by the insignificant results for the majority of these effects. }
	\item[rGPPM-c] This is another restriction of the GPPM, where we retain the cyclic calendar time component of the GPPM as well as the recency and lifetime components, but replace the other two unpredictable calendar time components with a constant. In theory, the full GPPM should behave similarly to this model in the holdout period, as the calendar time effects revert to a constant. In practice, however, it is possible that the lifetime component is estimated with greater precision when the calendar time effects are not included, and thus forecasting performance of the rGPPM-c may improve over the full model. 
\end{description}

\subsection{Predictive Performance}

Table~\ref{tab:fit} shows the predictive performance of the above benchmark models and that of the GPPM. The table reports the mean absolute percentage error (MAPE) and the root mean squared error (RMSE) for the calibration and holdout datasets of the two games. We see from the table that both log-logistic hazard models perform poorly. These models are estimated using the whole range of the data (i.e., with the calibration and holdout data combined). We see that even when using the whole range of the data for estimation, the predictive fit of these models in the holdout data is very poor.  Given this poor performance using the full data, we do not estimate these models separately on the partial calibration data. 

\begin{figure}[p!]
\centering
\makebox[\textwidth][c]{\includegraphics[scale=0.75]{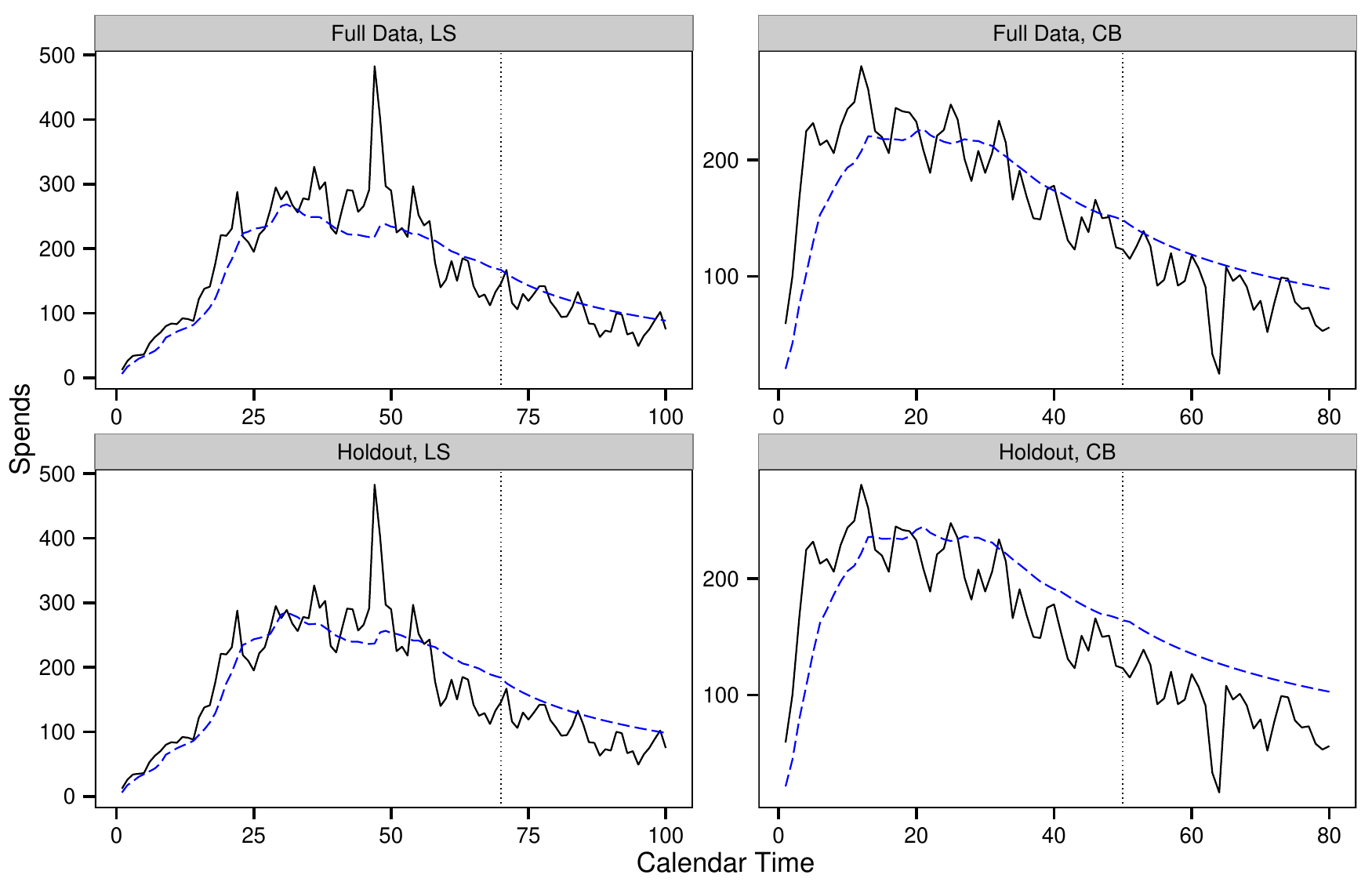}}
\caption{The BGNBD fit to both datasets; we fit it twice, once using all of the data (top row), and once saving the last 30 days for holdout (bottom row).}
\label{fig:bgnbdfit}
\end{figure}

Of primary interest to us is the comparison with the customer base analysis models. We see that the fit of the BGNBD is much better than that of the hazard models. The BGNBD fits, both fully in-sample and with training and holdout data, are presented in Figure~\ref{fig:bgnbdfit}. The panels associated with LS portray a fit that is similar to the fit of the muted GPPM in the bottom left panel of Figure~\ref{fig:ls_fit_decomp}, where all calendar time components are muted. This provides initial evidence that the patterns captured by the GPPM on the recency and lifetime scales are the same patterns that are captured by the BGNBD and other buy-till-you-die models. 

\begin{figure}[p!]
\centering
\makebox[\textwidth][c]{\includegraphics[scale=0.75]{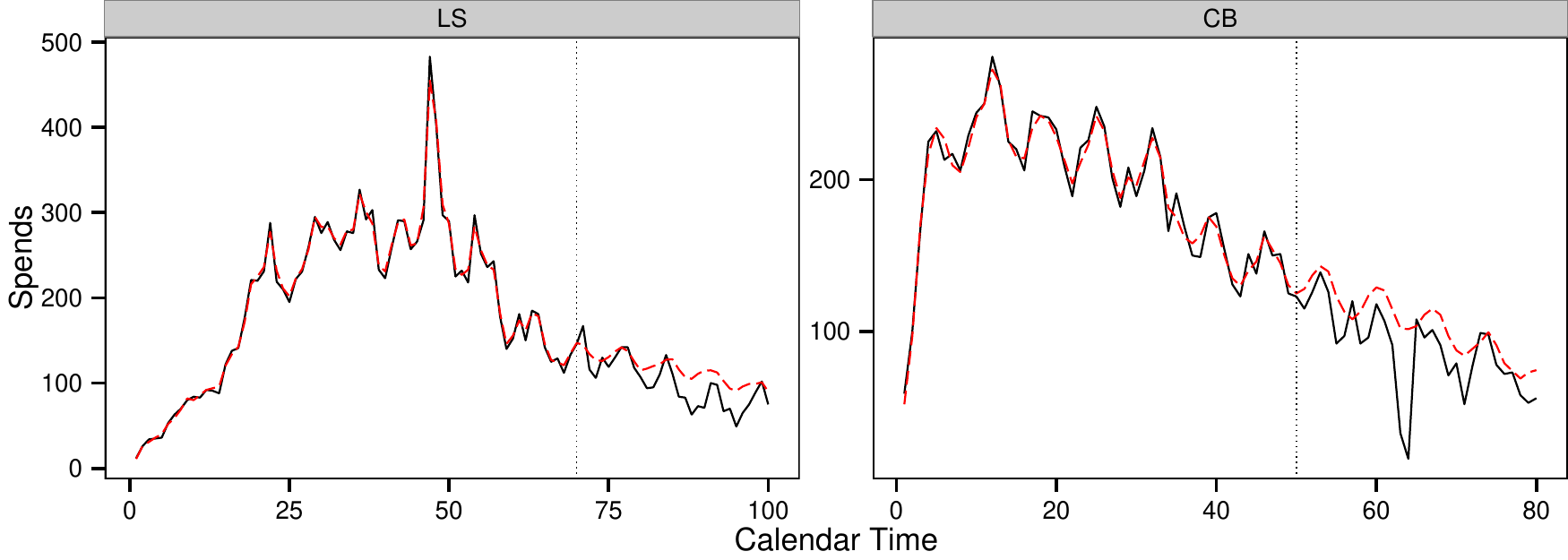} }
\caption{Life Simulator and City Builder out-of-sample predictions. The GPPM fit is dashed, while the data is solid.}
\label{fig:fore_fit}
\end{figure}

Crucially, the fit of the full GPPM and its reduced variants is significantly better than all the benchmarks, both in and out-of-sample. The forecast fit of the full GPPM is plotted in Figure~\ref{fig:fore_fit}. We can see that out-of-sample the full GPPM captures the mean and direction of daily spending, as well as cyclic patterns. Table~\ref{tab:fit} shows that both of the reduced models (rGPPM and rGPPM-c) have a worse in-sample fit when compared the full model. However, both reduced variants perform slightly better out-of-sample, as hypothesized. The difference is not huge, but notable. Of the three, the rGPPM-c fits best out-of-sample, again as hypothesized, as it contains all of the highly predictable components, but nothing else. This is consistent with our story in Section 3.3.1 and with the top-right and bottom-left panels in Figure~\ref{fig:ls_fit_decomp}. Recency and lifetime alone are able to capture a substantial portion of the dynamics, and the predictable cyclic effect explains still more variability. The GPPM framework can capture a very flexible set of recency and lifetime patterns, which allows it to outperform models like the BGNBD.

\subsection{Simulation Studies}

The GPPM provides a natural generalization of customer base analysis models that rely solely on recency and lifetime, such as the BGNBD. While the GPPM does not explicitly account for customer death, it does so asymptotically by allowing the probability of purchase to go to zero via the lifetime and recency effects. To explore this link deeper, we ran a series of simulation studies, testing in which cases the GPPM is able to capture BGNBD data, and vice versa. 

The rGPPM is the direct analogue of the BGNBD within the GPPM framework, as it is also restricted to recency and lifetime effects. We hypothesize that the dynamic spending patterns that are captured by the BGNBD can also be captured by the rGPPM; however, the BGNBD will have a difficult time fitting data generated by the full GPPM, depending on the strength of calendar time effects present. To test these two hypotheses, we first see how the rGPPM does at fitting data generated by the BGNBD model. Then we do the reverse and estimate the BGNBD on data from GPPMs that vary on the strength and nature of the calendar time effects. 

\paragraph{BGNBD Data, rGPPM Fit} 

\begin{figure}[p!]
\centering
\includegraphics[scale=0.75]{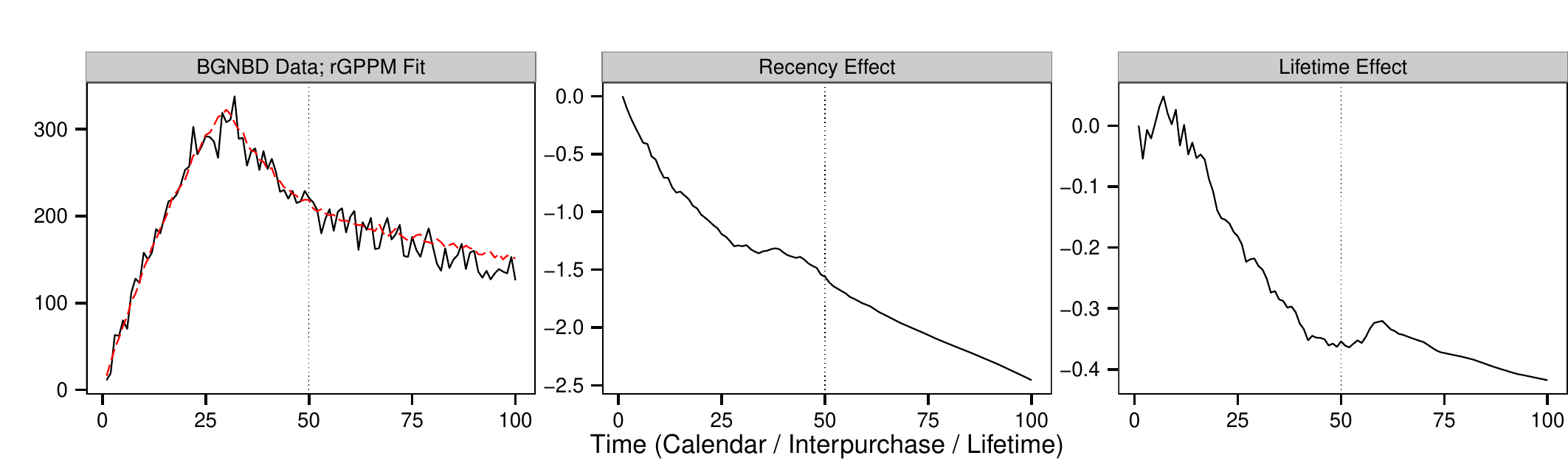}
\caption{Simulated data from the BGNBD model with the parameters from the original FHL paper, fit by the rGPPM model. At left, we see the fit; at right, we see the two latent function estimates, $\alpha_R(r_{it})$ and $\alpha_\mathcal{L}(\ell_{it})$ respectively, with their forecasts.}
\label{fig:fhl}
\end{figure}

If the recency and lifetime components of the GPPM do capture the dynamic patterns inherent in the BGNBD, then the rGPPM should be able to do well on data generated from the BGNBD.  To see this, we generate data from 10,000 spenders across 30 first spend dates, similar to our real data. We simulate spending over 100 days according to a BGNBD model, and then fit the rGPPM on the first 50 days of simulated data, and forecast the activity on days 51 to 100.  As our main example, we use the estimated BGNBD parameters ($r=0.243$, $\alpha=4.414$, $a=0.793$, $b=2.426$) from the original BGNBD paper (Fader, Hardie, and Lee, 2010, subsequently FHL). We also used many combinations of randomly generated parameters to test robustness. The simulated fit for the rGPPM on the FHL data is shown in the left panel of Figure~\ref{fig:fhl}.\footnote{Plots showing the fit for the 20 simulations are available from the authors, on request.} We see that the rGPPM model fits quite well, both in the training data and in holdout. The fit statistics for all of the simulations are summarized in Table~\ref{tab:bgnbd_vs_gppm}. The good fit offers substantial evidence to our claim that the GPPM nests these traditional probability models. 

While the rGPPM captures the variation along the two time scales on which the BGNBD operates, the nature of the analysis in the two models is fundamentally different. The BGNBD  represents the lifetime effects of a strict active/inactive nature, whereas the rGPPM allows for a gradual decrease in spending rates over the customer lifetime. We can understand the dynamic patterns implied by the BGNBD by looking at these through the lens of the rGPPM. In particular, we can look at the implied latent curves, $\alpha_{R}(r_{it})$ and $\alpha_\mathcal{L}(\ell_{it})$, in the right two panels of Figure~\ref{fig:fhl}. We see that the BGNBD with the FHL parameters implies a strong recency effect, and a weaker and slightly non-monotonic lifetime effect that appears to flatten out around day 50. We believe that a detailed investigation of the correspondence between these two models is likely to yield further interesting insights.

\paragraph{GPPM Data, BGNBD Fit}

We also study the reverse situation and examine the performance of the BGNBD on data generated from the GPPM. We show that BGNBD is not able to fit such data very well, especially in the presence of calendar time dynamics. Specifically, we use three levels of the day of the week effect --- none (\texttt{Nocyc}), weak (\texttt{Weakcyc}), and strong (\texttt{Strongcyc}) --- and  three kinds of non-cyclic calendar time effects: none (\texttt{Nocal}), a long-run peak similar to the general holiday season bump seen above (\texttt{Peakcal}), and a nonlinear decreasing trend across the whole time period (\texttt{NonlinDeccal}).\footnote{The cyclic effect was set as $\alpha_w(t) = \theta \sin(2 \pi t / 7)$, where $\theta=0$, for no cyclic effect, $\theta=0.15$, for the weak effect, and $\theta=0.4$, for the strong effect. For the calendar time effects, the non-linear decreasing calendar time trend is given by $\alpha_{T}^\ell(t) = -0.2 t^{0.3}$; the peak effect is given by the piecewise function: $ \alpha_{T}^\ell(t) = 0$, when $t \le 20$; $ \alpha_{T}^\ell(t) = 0.5(t-20)$, when $ t \in [21,40] $; $ \alpha_{T}^\ell(t) = 0.1(50-t) $, when $ t \in [41,50]$ and $ \alpha_{T}^\ell(t) = 0$, when $t >50$. }

\begin{figure}[p!]
\centering
\includegraphics[scale=0.75]{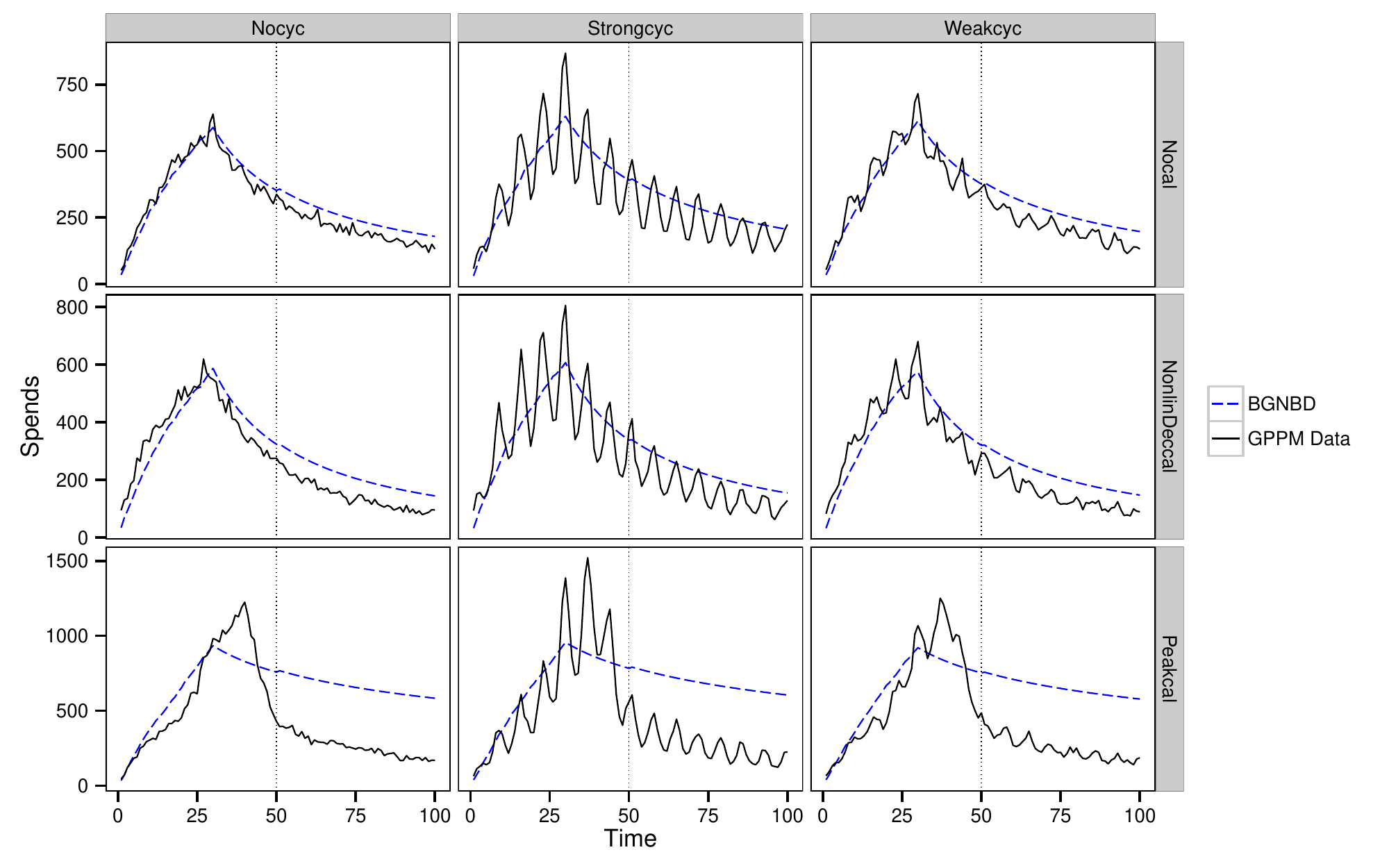}
\caption{The BGNBD fit on various types of data drawn from the GPPM: \texttt{Nocyc}, \texttt{Strongcyc}, and \texttt{Weakcyc} indicate no, strong, and weak cyclic (day of the week) effects respectively; \texttt{Nocal} indicates no calendar time dynamics, \texttt{NonlinDeccal} indicates a non-linear decreasing long-run calendar time process, and \texttt{Peakcal} indicates a calendar time process that is flat but with a peak during the calibration period.}
\label{fig:gppdata_bgnbdfit}
\end{figure}

\begin{figure}[p!]
\centering
\includegraphics[scale=0.75]{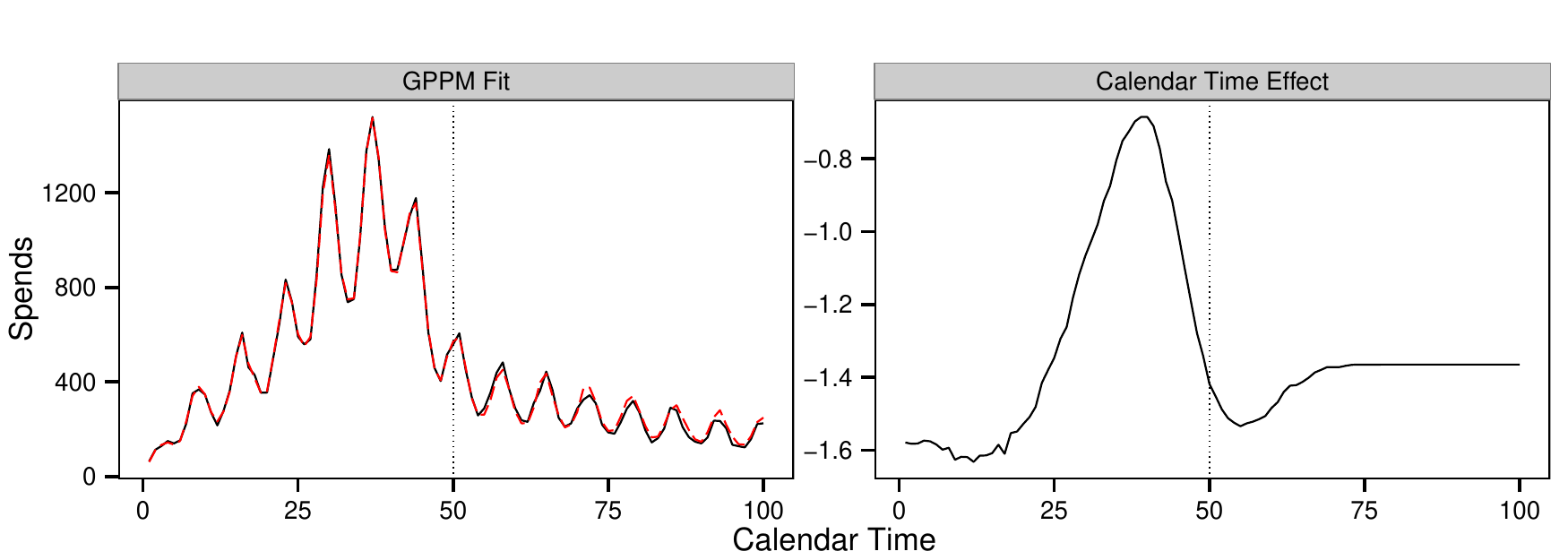}
\caption{The GPPM fit and forecast on the \texttt{Strongcyc/Peakcal} simulated data, together with the estimated calendar time effect.}
\label{fig:gppdata_gppfit}
\end{figure}

Figure~\ref{fig:gppdata_bgnbdfit} and Table~\ref{tab:bgnbd_vs_gppm} show the results from these simulations. We see that BGNBD fits the mean of the curve in the presence of a cyclic effect. We also see that the BGNBD generally does well in the cases where there is no short or long-run calendar variation, underpredicts in the beginning and then overpredicts in the end when there is a decreasing calendar time effect, and fails significantly at capturing the peak effect.  We see in the \texttt{Peakcal} case (last row of Figure~\ref{fig:gppdata_bgnbdfit}) that the BGNBD attributes the peak to higher rates of spending, and then dramatically overestimates future spending.  The GPPM does not fall prey to this same bias because of its ability to separate out calendar time effects. To emphasize this, we see the GPPM fit to the worst case (\texttt{Strongcyc/Peakcal}), together with the estimated calendar time effect, in Figure~\ref{fig:gppdata_gppfit}.

\section{Conclusion}
In this paper, we developed a highly flexible model-based approach for understanding and predicting spending dynamics across customer bases. Our Gaussian process propensity model employs Bayesian nonparametric Gaussian process priors to decompose a latent propensity to spend process into components that vary along calendar time, interpurchase time, and customer lifetime dimensions. This additive structure yields easily interpretable model outputs and fits customer spending data well.

We showed that the GPPM identifies the latent dynamic patterns in the data via a principled probabilistic framework that reliably separates signal from noise. It offers a number of outputs that are of considerable value to managers. First, the GPPM generates a visual dashboard of latent functions. These model-based visualizations are easy to comprehend, even by managers who may lack sophisticated statistical skills.  Second, we demonstrated that the GPPM is capable of automatically capturing the effect of events that may be of interest to managers. This is particularly useful in settings where certain events may escape their notice. In these situations, the GPPM is able to flag such events and bring them to the awareness of decision makers. More importantly, the nonparametric nature of the GPPM allows it to flexibly model the nature and duration of the impact of events (either known or unknown, a priori), without the need to represent these explicitly via covariates. Third, we also showed that the GPPM can be updated dynamically to  capture ``real-time"  dynamics as they unfold and is thus useful to managers who often rely on dashboards to uncover shifting patterns of purchasing activity in the customer base. All of the above advantages of the GPPM make it ideal for decision contexts involving multiple products and information asymmetries. 

We demonstrated these benefits of the GPPM on two data sets that characterize purchasing activity within mobile games. Apart from illustrating these benefits, we also showed that the GPPM outperforms traditional customer base analysis models in terms of predictive performance, both in-sample and out-of-sample. The predictive superiority of the GPPM stems from the fact that it implicitly nests traditional customer base analysis models, including hazard models with time-varying covariates and the class of buy-till-you-die models (BTYD), such as the Pareto-NBD, BGBB, and BGNBD. 

While the paper showcases the many benefits of our framework, it is also important to acknowledge some limitations.  Our assumption that all customers follow the same dynamic process, apart from a level shift caused by the random effect, is limiting. Further, the assumption of only two characteristic length-scales for the calendar time process may not hold in all situations. Future research can explore the impact of richer kernel structures on the ability to forecast long-run trends, and to separate short-term and long-run effects. Another limitation stems from the need for a large amount of data for estimation. For instance, we used complete time series data to estimate the model. Some of the benchmark models, in particular, the BGNBD and the Pareto-NBD, use only two sufficient statistics per customer. Although we think that the data needed for the GPPM are widely available in practice, the data and computational demands are much steeper than for competing models.

To conclude, we believe the GPPM addresses a fundamental need of modern marketing managers for flexible, model-based decision support systems. Our work contributes to the literatures on customer base analysis and decision support, while simultaneously introducing a new Bayesian nonparametric approach to the marketing literature. Dashboards and marketing analytics systems are likely to become even more important in the future, given the increasing complexity of modern data-rich environments. As dashboards increase in relevance, we believe that managers will welcome further academic research in this domain.

\singlespacing

\bibliography{gppmbib}

\begin{thebibliography}{}

\bibitem[Ascarza and Hardie, 2013]{Ascarza2013}
Ascarza, E. and Hardie, B. G.~S. (2013).
\newblock {A Joint Model of Usage and Churn in Contractual Settings}.
\newblock {\em Marketing Science}, 32:570--590.

\bibitem[Chatfield and Goodhard, 1973]{Chatfield1973}
Chatfield, C. and Goodhard, G.~J. (1973).
\newblock {A consumer purchasing model with Erlang inter-purchase times}.
\newblock {\em Journal of the American Statistical Association},
  68(344):828--835.

\bibitem[Duvenaud et~al., 2013]{Duvenaud2013}
Duvenaud, D., Lloyd, J., Grosse, R., Tenenbaum, J., and Ghahramani, Z. (2013).
\newblock {Structure discovery in nonparametric regression through
  compositional kernel search}.
\newblock {\em Proceedings of the International Conference on Machine Learning
  (ICML)}, 30:1166--1174.

\bibitem[Fader et~al., 2005a]{Fader2005}
Fader, P., Hardie, B., and Lee, K.~L. (2005a).
\newblock {Counting Your Customers the Easy Way: An Alternative to the
  Pareto/NBD Model}.
\newblock {\em Marketing Science}, 24(2):275--284.

\bibitem[Fader et~al., 2005b]{Fader2005a}
Fader, P., Hardie, B., and Lee, K.~L. (2005b).
\newblock {RFM and CLV: Using Iso-Value Curves for Customer Base Analysis}.
\newblock {\em Journal of Marketing Research}, 42(4):415--430.

\bibitem[Fader et~al., 2010]{Fader2010}
Fader, P., Hardie, B., and Shang, J. (2010).
\newblock {Customer-Base Analysis in a Discrete-Time Noncontractual Setting}.
\newblock {\em Marketing Science}, 29(6):1086--1108.

\bibitem[Gelman and Rubin, 1992]{Gelman1992}
Gelman, A. and Rubin, D.~B. (1992).
\newblock {Inference from Iterative Simulation Using Multiple Sequences}.
\newblock {\em Statistical Science}, 7(4):457--511.

\bibitem[Genton, 2001]{Genton2001}
Genton, M.~G. (2001).
\newblock {Classes of Kernels for Machine Learning: A Statistics Perspective}.
\newblock {\em Journal of Machine Learning Research}, 2:299--312.

\bibitem[Gupta, 1991]{Gupta1991}
Gupta, S. (1991).
\newblock {Stochastic models of interpurchase time with time-dependent
  covariates}.
\newblock {\em Journal of Marketing Research}, 28:1--15.

\bibitem[Hanssens et~al., 2001]{Hanssens2001}
Hanssens, D.~M., Parsons, L.~J., and Schultz, R.~L. (2001).
\newblock {\em {Market Response Models: Econometric and Time Series Analysis}}.
\newblock Kluwer Academic Publishers, 2nd edition.

\bibitem[Helsen and Schmittlein, 1993]{Helsen1993}
Helsen, K. and Schmittlein, D.~C. (1993).
\newblock {Analyzing Duration Times in Marketing: Evidence for the
  Effectiveness of Hazard Rate Models}.
\newblock {\em Marketing Science}, 12(4):395--414.

\bibitem[Hoffman and Gelman, 2014]{Hoffman2011}
Hoffman, M. and Gelman, A. (2014).
\newblock {The no-U-turn sampler: Adaptively setting path lengths in
  Hamiltonian Monte Carlo}.
\newblock {\em Journal of Machine Learning Research}, 15:1351--1381.

\bibitem[Jain and Vilcassim, 1991]{Jain1991}
Jain, D.~C. and Vilcassim, N.~J. (1991).
\newblock {Investigating Household Purchase Timing Decisions: A Conditional
  Hazard Function Approach}.
\newblock {\em Marketing Science}, 10(1):1--23.

\bibitem[Kim et~al., 2007]{Kim2007}
Kim, J.~G., Menzefricke, U., and Feinberg, F.~M. (2007).
\newblock {Capturing Flexible Heterogeneous Utility Curves: A Bayesian Spline
  Approach}.
\newblock {\em Management Science}, 53(2):340--354.

\bibitem[Li and Ansari, 2014]{Li2014}
Li, Y. and Ansari, A. (2014).
\newblock {A Bayesian Semiparametric Approach for Endogeneity and Heterogeneity
  in Choice Models}.
\newblock {\em Management Science}, 60(5):1161--1179.

\bibitem[Morrison and Schmittlein, 1981]{Morrison1981}
Morrison, D.~G. and Schmittlein, D.~C. (1981).
\newblock {Predicting Future Random Events Based on Past Performance}.
\newblock {\em Management Science}, 27(9):1006--1023.

\bibitem[Neal, 2011]{Neal2011}
Neal, R.~M. (2011).
\newblock {MCMC using Hamiltonian dynamics}.
\newblock {\em Handbook of Markov Chain Monte Carlo}, pages 113--162.

\bibitem[Neslin et~al., 2006]{Neslin2006}
Neslin, S.~A., Gupta, S., Kamakura, W., Lu, J., and Mason, C.~H. (2006).
\newblock {Defection Detection: Measuring and Understanding the Predictive
  Accuracy of Customer Churn Models}.
\newblock {\em Journal of Marketing Research}, 43(2):204--211.

\bibitem[Pauwels et~al., 2009]{Pauwels2009}
Pauwels, K., Ambler, T., Clark, B.~H., LaPointe, P., Reibstein, D., Skiera, B.,
  Wierenga, B., and Wiesel, T. (2009).
\newblock {Dashboards as a Service: Why, What, How, and What Research Is
  Needed?}
\newblock {\em Journal of Service Research}, 12(2):175--189.

\bibitem[Rasmussen and Williams, 2006]{Rasmussen2006}
Rasmussen, E. and Williams, K.~I. (2006).
\newblock {\em {Gaussian Processes for Machine Learning‎}}.
\newblock MIT Press.

\bibitem[Roberts et~al., 2013]{Roberts2013}
Roberts, S., Osborne, M., Ebden, M., Reece, S., Gibson, N., and Aigrain, S.
  (2013).
\newblock {Gaussian processes for time-series modelling.}
\newblock {\em Philosophical transactions. Series A, Mathematical, physical,
  and engineering sciences}, 371(1984):20110550.

\bibitem[Rossi, 2013]{Rossi2013}
Rossi, P.~E. (2013).
\newblock {Bayesian Semi-parametric and Non-parametric Methods with
  Applications to Marketing and Micro-econometrics}.

\bibitem[Schmittlein et~al., 1987]{Schmittlein1987}
Schmittlein, D.~C., Morrison, D.~G., and Colombo, R. (1987).
\newblock {Counting Your Customers: Who-Are They and What Will They Do Next?}
\newblock {\em Management Science}, 33(1):1--24.

\bibitem[Schweidel and Knox, 2013]{Schweidel2013}
Schweidel, D.~A. and Knox, G. (2013).
\newblock {Incorporating Direct Marketing Activity into Latent Attrition
  Models}.
\newblock {\em Marketing Science}, 32(3):471--487.

\bibitem[Seetharaman and Chintagunta, 2003]{Seetharaman2003}
Seetharaman, P.~B. and Chintagunta, P.~K. (2003).
\newblock {The Proportional Hazard Model for Purchase Timing: A Comparison of
  Alternative Specifications}.
\newblock {\em Journal of Business \& Economic Statistics}, 21(3):368--382.

\bibitem[Wedel and Zhang, 2004]{Wedel2004}
Wedel, M. and Zhang, J. (2004).
\newblock {Analyzing Brand Competition Across Subcategories}.
\newblock {\em Journal of Marketing Research}, 41:448--456.

\end{thebibliography}

\newpage

\clearpage

\begin{table}[p!]
	\centering
	\caption{Posterior medians for kernel hyperparameters}
	\begin{tabular}{lrr@{\hskip 1cm}rr}
		\toprule
		& \multicolumn{2}{c}{Life Simulator} & \multicolumn{2}{c}{City Builder} \\
		\midrule
		& Amplitude & Length-scale & Amplitude & Length-scale \\ 
		\midrule
		Long-run & 0.026 & 104.735 & 0.075 & 94.890 \\  
		Short-run & 0.023 & 2.219 & 0.018 & 69.396 \\ 
		Recency & 0.002 & 114.197 & 0.010 & 110.880 \\  
		Lifetime & 0.002 & 82.613 & 0.053 & 144.595 \\ 
		Cyclic & 0.994 & 49.906 & 0.800 & 42.560 \\ 
		\bottomrule
	\end{tabular} 
	\label{tab:hyperpars}
\end{table}

\begin{table}[p!]
	\centering
	\caption{Other parameters, posterior summaries}
	\begin{tabular}{clrrr@{\hskip 1cm}rrr}
		\toprule
		& & \multicolumn{3}{c}{Life Simulator} & \multicolumn{3}{c}{City Builder} \\
		& & Median & 2.5\% & 97.5\%  & Median & 2.5\% & 97.5\%\\ 
		\midrule
		\multirow{5}{1.25in}{Mean function hyperparameters} & $\mu$ & -1.677 & -1.882 & -1.435 & -1.793 & -2.185 & -1.381 \\ 
		& $\lambda^r_1$ & 0.571 & 0.525 & 0.655 & 0.134 & 0.076 & 0.215 \\ 
		& $\lambda^{cl}_1$ & 0.368 & 0.258 & 0.447 & 0.772 & 0.615 & 0.896 \\ 
		& $\lambda^r_2$ & 0.503 & 0.466 & 0.526 & 0.715 & 0.580 & 0.855 \\ 
		& $\lambda^{cl}_2$ & 0.250 & 0.184 & 0.348 & 0.351 & 0.300 & 0.407 \\ 
		\midrule
		\multirow{3}{1.25in}{Random effect variances} & $\sigma^2_{\delta}$ & 0.676 & 0.622 & 0.754 & 0.913 & 0.832 & 0.984 \\ 
		& $\sigma^2_{\mathrm{First~Spend}}$ & 0.079 & 0.029 & 0.111 & 0.037 & 0.006 & 0.089 \\ 
		& $\sigma^2_{\mathrm{Install}}$ & 0.028 & 0.006 & 0.079 & 0.053 & 0.003 & 0.117 \\ 
		\midrule
		\multirow{9}{1.25in}{Acquisition channel fixed effects} & Ad 1 & 0.000 & 0.000 & 0.000 & 0.000 & 0.000 & 0.000 \\  
		& Ad 2 & 0.082 & -0.199 & 0.542 & -0.322 & -0.841 & 0.049 \\ 
		& Ad 3 & 0.191 & -0.114 & 0.455  & - & - & - \\
		& Ad 4 & - & - & - & -0.462 & -0.780 & -0.156 \\
		& Ad 5 & - & - & - & 0.044 & -0.242 & 0.343 \\ 
		& Social 1 & 0.282 & 0.083 & 0.523 & 0.143 & -0.116 & 0.430 \\ 
		& Social 2 & -0.066 & -0.495 & 0.303 & -0.101 & -0.471 & 0.264 \\ 
		& Company & 0.468 & 0.245 & 0.701 & 0.113 & -0.148 & 0.426 \\ 
		& Organic & 0.151 & -0.001 & 0.353 & -0.097 & -0.303 & 0.132 \\ 
		\bottomrule
	\end{tabular}
	\label{tab:pars}
\end{table}

\begin{table}[p!]
\centering
\begin{tabular}{llrrr@{\hskip 2em}rrr}
  \toprule
  & & \multicolumn{3}{c}{Life Simulator} & \multicolumn{3}{c}{City Builder} \\
  & & Overall & Training & Holdout &  Overall & Training & Holdout \\ 

  \midrule
  \multirow{2}{*}{GPPM} & MAPE & 0.09 & 0.03 & 0.24 & 0.18 & 0.03 & 0.42 \\ 
  & RMSE & 13.88 & 7.05 & 22.93 & 16.79 & 7.08 & 25.86 \\ 
  \midrule
  \multirow{2}{*}{rGPPM} & MAPE & 0.14 & 0.14 & 0.15 & 0.20 & 0.10 & 0.37 \\ 
  & RMSE & 38.01 & 43.64 & 19.31 & 23.15 & 21.64 & 25.46 \\ 
  \midrule
  \multirow{2}{*}{rGPPM-c} & MAPE & 0.12 & 0.11 & 0.14  & 0.16 & 0.06 & 0.32  \\ 
  & RMSE & 30.06 & 34.39 & 15.88 & 16.96 & 13.06 & 21.96 \\ 
  \midrule
  \multirow{2}{*}{Log-Logistic*} & MAPE & 0.42 & 0.31 & 0.66 & 0.41 & 0.19 & 0.76\\ 
  & RMSE & 68.26 & 71.82 & 59.10 &  46.76 & 47.00 & 46.36 \\ 
  \midrule
  \multirow{2}{*}{L-L Covs*}  & MAPE & 0.27 & 0.19 & 0.47 & 0.28 & 0.15 & 0.49 \\ 
  & RMSE & 62.83 & 66.98 & 51.45 & 36.41 & 32.70 & 42.06  \\ 
  \midrule
  \multirow{2}{*}{BGNBD} &  MAPE & 0.25 & 0.20 & 0.36 & 0.40 & 0.20 & 0.73  \\ 
  & RMSE & 46.09 & 50.34 & 34.18 & 42.70 & 41.31 & 44.94\\ 
  \bottomrule
\end{tabular} 
\caption{Fit statistics for the LS and CB forecasts across the GPPM and benchmark models. Models indicated by * were estimated using the whole range of data (that is, with no data left for holdout), although we still break down the fit by training and holdout. The fit statistics for the rGPPM and rGPPM-c do not change substantially with the inclusion/exclusion of observed heterogeneity.}
\label{tab:fit}
\end{table}

\begin{table}[p!]
\centering
\begin{tabular}{lrrr@{\hskip 2em}rrr}
  \toprule
   & \multicolumn{3}{c}{FHL Parameters} & \multicolumn{3}{c}{Random Simulations} \\
 & Overall & Training & Holdout  & Overall & Training & Holdout \\
  \midrule
MAPE & 0.08 & 0.07 & 0.09  & 0.07 & 0.04 & 0.10 \\ 
  RMSE & 15.44 & 13.10 & 17.48 & 37.28 & 31.31 & 41.83  \\
\end{tabular}
\begin{tabular}{lrrr@{\hskip 2em}rrr}
  \midrule \midrule
  & \multicolumn{3}{c}{All Simulations} & \multicolumn{3}{c}{No Calendar Dynamics}  \\
 & Overall & Training & Holdout &  Overall & Training & Holdout \\  
  \midrule
MAPE & 0.54 & 0.21 & 0.87 & 0.22 & 0.15 & 0.29 \\ 
  RMSE & 153.76 & 110.76 & 181.71 & 64.26 & 65.67 & 61.35 \\
  \bottomrule
\end{tabular}
\caption{Fit summaries for the two simulation studies. In the top panel, we present results for the rGPPM model estimated on data generated from the BGNBD. At left are the fit statistics for the data simulated from the FHL parameters. At right is the summary across 20 randomly drawn cases, where mean fit statistics are reported. In the bottom panel, we present average fit statistics for the BGNBD estimated on data simulated from GPPM. At left are the statistics across all the simulations; at right are the statistics in only the \texttt{Nocal} condition. We argue that the BGNBD is equivalent to a subclass of GPPM models with no calendar time dynamics; the superior performance of the BGNBD in this condition than across all the conditions supports this hypothesis.}
\label{tab:bgnbd_vs_gppm}
\end{table}

\end{document}